\documentclass{article}
\PassOptionsToPackage{numbers,compress}{natbib}
\usepackage[preprint]{neurips_2026}
\usepackage[utf8]{inputenc}
\usepackage[T1]{fontenc}
\usepackage{hyperref}
\usepackage{url}
\usepackage{booktabs}
\usepackage{amsfonts}
\usepackage{nicefrac}
\usepackage{microtype}
\usepackage{xcolor}
\usepackage{graphicx}
\usepackage{makecell}
\usepackage{subcaption}
\usepackage{tabularx}
\usepackage{array}
\usepackage{float}
\usepackage{amsmath, amssymb, mathtools, amsthm}
\usepackage{pifont}
\usepackage{enumitem}

\usepackage[capitalize,noabbrev]{cleveref}
\usepackage{fancyhdr}
\fancypagestyle{fermilabpub}{%
  \fancyhf{}%
  \fancyhead[R]{FERMILAB-PUB-26-0347-CMS-LDRD}%
  \fancyfoot[C]{\thepage}%
}
\usepackage{tikz}
\usetikzlibrary{shapes, arrows.meta, positioning, calc, fit, backgrounds, shadows.blur}
\definecolor{myblue}{RGB}{70, 130, 180}
\definecolor{myred}{RGB}{220, 20, 60}
\definecolor{mygreen}{RGB}{34, 139, 34}
\definecolor{myorange}{RGB}{255, 140, 0}
\usepackage{amsmath,amsfonts,bm}

\def\eqref#1{equation~\ref{#1}}
\def\1{\bm{1}}

\def\rr{{\textnormal{r}}}

\DeclareMathAlphabet{\mathsfit}{\encodingdefault}{\sfdefault}{m}{sl}
\SetMathAlphabet{\mathsfit}{bold}{\encodingdefault}{\sfdefault}{bx}{n}

\def\tA{{\tens{A}}}
\def\tB{{\tens{B}}}
\def\tC{{\tens{C}}}

\def\tR{{\tens{R}}}

\def\tW{{\tens{W}}}

\def\sA{{\mathbb{A}}}
\def\sB{{\mathbb{B}}}
\def\sC{{\mathbb{C}}}

\def\sR{{\mathbb{R}}}

\def\sW{{\mathbb{W}}}

\newcommand{\R}{\mathbb{R}}

\newcommand{\kT}{k_{\mathrm{T}}}
\newcommand{\pT}{p_{\mathrm{T}}}
\newcommand{\gain}[1]{\textcolor{mygreen}{\scriptsize $\uparrow$ #1}}

\title{Patch Hierarchical Attention Transformer for Efficient Particle Jet Tagging}

\author{%
  Aaron Wang\thanks{Equal contribution.} \\
  University of Illinois Chicago \\
  Chicago, IL 60607, USA \\
  \texttt{aaronw5@uic.edu} \\
  \And
  Zihan Zhao\footnotemark[1] \\
  University of California San Diego \\
  La Jolla, CA 92093, USA \\
  \texttt{ziz078@ucsd.edu} \\
  \And
  Alan Xia\footnotemark[1] \\
  University of California San Diego \\
  La Jolla, CA 92093, USA \\
  \AND
  Chang Sun \\
  California Institute of Technology  \\
  Pasadena, CA 91125, USA \\
  \AND
  Abhijith Gandrakota \\
  Fermi National Accelerator Laboratory \\
  Batavia, IL 60510, USA \\
  \And
  Jennifer Ngadiuba \\
  Fermi National Accelerator Laboratory \\
  Batavia, IL 60510, USA \\
  \And
  Richard Cavanaugh \\
  University of Illinois Chicago \\
  Chicago, IL 60607, USA \\
  \And
  Javier Duarte \\
  University of California San Diego \\
  La Jolla, CA 92093, USA \\
}
\begin{document}
\maketitle
\thispagestyle{fermilabpub}

\begin{abstract}
Real-time jet tagging is critical for identifying short-lived particle decays in the high-throughput detectors of the Large Hadron Collider, where real-time trigger systems responsible for deciding which collision events to store impose strict latency and accuracy constraints. While transformer architectures achieve the highest jet tagging accuracy when compute is unconstrained, their quadratic self-attention cost makes inference restrictive on trigger budget. Existing efficient variants reduce the computational cost, but hinder the classification performance. To address this limitation, we introduce the Patch Hierarchical Attention Transformer (PHAT-JeT), which combines two mechanisms: a physics-inspired geometric message-passing module that encodes local detector-plane structure, and a hierarchical patch-based attention scheme that computes exact attention within small particle groups while preserving global context through lightweight patch-token communication. Within a restricted budget, PHAT-JeT achieves state-of-the-art accuracy and background rejection among all resource-constrained jet tagging models on four benchmarks (\textsc{hls4ml}, JetClass, Top Tagging, and Quark--Gluon). Our code is available at \href{https://anonymous.4open.science/r/PHAT-JeT-540B/README.md}{anonymous.4open.science/r/PHAT-JeT-540B}.
\end{abstract}

\section{Introduction}
\label{sec:intro}
\begin{figure}[!ht]
\centering
\begin{subfigure}[b]{0.49\linewidth}
    \centering
    \resizebox{\linewidth}{!}{
        \begin{tikzpicture}[line cap=round, line join=round]
        \usetikzlibrary{arrows.meta}
        \definecolor{qgcolor}{RGB}{130, 70, 190}
        \definecolor{ecal}{RGB}{0, 160, 170}
        \definecolor{hcal}{RGB}{240, 170, 40}
        \definecolor{track}{RGB}{90, 100, 130}
        \begin{scope}[shift={(0,0)}]
        \def\R{2.40}
        \def\Rout{2.62}
        \fill[black!4] (0,0) circle (\Rout);
        \draw[thick, black!30] (0,0) circle (\Rout);
        \fill[white, opacity=0.70] (0,0) circle (\R);
        \foreach \rr in {0.55, 1.05, 1.55, 2.05} { \draw[black!9] (0,0) circle (\rr); }
        \foreach \a in {0,12,...,168} { \draw[black!7] (0,0) -- (\a:\R); }
        \draw[very thick, -{Latex[length=4mm]}, black!55] (-3.6,0) -- (-0.35,0);
        \draw[very thick, -{Latex[length=4mm]}, black!55] ( 3.6,0) -- ( 0.35,0);
        \node[black!55] at (-2.85,-0.35) {\large $p$};
        \node[black!55] at ( 2.85,-0.35) {\large $p$};
        \fill[black!65] (0,0) circle (2.2pt);
        \fill[black!18] (0,0) circle (6.6pt);
        \foreach \a/\rad/\cx/\cy in {
          -85/2.10/ 0.25/-0.70, -72/2.25/ 0.28/-0.65, -60/2.35/ 0.30/-0.55, -48/2.20/ 0.32/-0.45,
          -36/2.30/ 0.35/-0.35, -24/2.15/ 0.36/-0.25, -12/2.35/ 0.38/-0.15,    0/2.20/ 0.38/ 0.00,
           12/2.35/ 0.36/ 0.15,  24/2.25/ 0.34/ 0.30,  36/2.35/ 0.30/ 0.45,  48/2.15/ 0.24/ 0.60,
           60/2.30/ 0.15/ 0.70,  72/2.10/ 0.05/ 0.75,  84/2.25/-0.05/ 0.75,  96/2.10/-0.15/ 0.72,
          108/2.15/-0.26/ 0.65, 120/2.05/-0.36/ 0.55, 132/2.00/-0.45/ 0.42, 144/1.95/-0.52/ 0.28,
          156/1.90/-0.60/ 0.12, 168/1.85/-0.64/-0.05
        }{
          \path (0,0) ++(\a:\rad) coordinate (T);
          \draw[black!45, line width=0.8pt, opacity=0.20]
            (0,0) .. controls +(\cx,\cy) and +(-0.15,0.10) .. (T);
        }
        \foreach \a/\rad/\col/\cx/\cy in {
           44/2.25/hcal/ 0.26/ 0.55,
          122/2.00/ecal/-0.30/ 0.58,
          -30/2.20/hcal/ 0.35/-0.35,
          -62/2.30/track/0.28/-0.62
        }{
          \edef\tmpcol{\col}%
          \path (0,0) ++(\a:\rad) coordinate (TC);
          \expandafter\draw\expandafter[\tmpcol!75, line width=0.9pt, opacity=0.18]
            (0,0) .. controls +(\cx,\cy) and +(-0.15,0.10) .. (TC);
        }
        \foreach \a/\h/\col in {
          205/0.20/qgcolor, 214/0.12/qgcolor, 222/0.16/qgcolor, 232/0.26/qgcolor, 240/0.14/qgcolor,
          252/0.14/hcal,    260/0.18/hcal,    270/0.22/hcal,    282/0.12/hcal,
          296/0.14/ecal,    306/0.18/ecal,    316/0.12/ecal,
           18/0.10/hcal,     30/0.12/ecal,     42/0.10/hcal,
          142/0.10/black!60, 152/0.12/hcal,    162/0.10/ecal
        }{
          \edef\tmpcol{\col}%
          \begin{scope}[shift={(\a:\R)}, rotate=\a]
            \expandafter\fill\expandafter[\tmpcol, opacity=0.70] (-0.06,0) rectangle (0.06,\h);
          \end{scope}
        }
        \def\JoneA{56}
        \def\JoneB{88}
        \fill[qgcolor, opacity=0.22] (0,0) -- (\JoneA:\R) arc (\JoneA:\JoneB:\R) -- cycle;
        \draw[qgcolor!80, line width=1.35pt, opacity=0.75] (0,0) -- (\JoneA:\R);
        \draw[qgcolor!80, line width=1.35pt, opacity=0.75] (0,0) -- (\JoneB:\R);
        \draw[qgcolor!80, line width=1.35pt, opacity=0.65] (\JoneA:\R) arc (\JoneA:\JoneB:\R);
        \def\JoneAxis{73}
        \coordinate (SPLIT1) at (\JoneAxis:0.95);
        \draw[qgcolor!70, line width=0.9pt, opacity=0.28] (0,0) -- (SPLIT1);
        \fill[qgcolor!60, opacity=0.50] (SPLIT1) circle (2.0pt);
        \foreach \k in {1,...,5} {
          \pgfmathsetmacro{\ang}{\JoneAxis - 11 + \k*4.2}
          \path (0,0) ++(\ang:1.65) coordinate (F);
          \draw[qgcolor!80, thin, opacity=0.13] (0,0) .. controls +(0.18,0.55) and +(-0.10,0.10) .. (F);
        }
        \coordinate (Oa) at ($(SPLIT1) + (-0.07,0.03)$);
        \coordinate (Ob) at ($(SPLIT1) + ( 0.02,0.06)$);
        \coordinate (Oc) at ($(SPLIT1) + ( 0.07,-0.02)$);
        \def\sR{1.62} \def\sW{6.0} \def\sA{62} \def\sB{74} \def\sC{86}
        \draw[black!35, thin, opacity=0.25] (SPLIT1) .. controls +(0.05,0.10) and +(-0.05,0.02) .. (Oa);
        \draw[black!35, thin, opacity=0.25] (SPLIT1) .. controls +(0.05,0.10) and +(-0.05,0.02) .. (Ob);
        \draw[black!35, thin, opacity=0.25] (SPLIT1) .. controls +(0.05,0.10) and +(-0.05,0.02) .. (Oc);
        \fill[ecal,  opacity=0.24] (Oa) -- ++(\sA-\sW:\sR) arc (\sA-\sW:\sA+\sW:\sR) -- cycle;
        \draw[ecal!85,  line width=1.15pt, opacity=0.88] (Oa) -- ++(\sA-\sW:\sR);
        \draw[ecal!85,  line width=1.15pt, opacity=0.88] (Oa) -- ++(\sA+\sW:\sR);
        \draw[ecal!85,  line width=1.15pt, opacity=0.82] (Oa) ++(\sA-\sW:\sR) arc (\sA-\sW:\sA+\sW:\sR);
        \fill[track, opacity=0.24] (Ob) -- ++(\sB-\sW:\sR) arc (\sB-\sW:\sB+\sW:\sR) -- cycle;
        \draw[track!85, line width=1.15pt, opacity=0.88] (Ob) -- ++(\sB-\sW:\sR);
        \draw[track!85, line width=1.15pt, opacity=0.88] (Ob) -- ++(\sB+\sW:\sR);
        \draw[track!85, line width=1.15pt, opacity=0.82] (Ob) ++(\sB-\sW:\sR) arc (\sB-\sW:\sB+\sW:\sR);
        \fill[hcal,  opacity=0.24] (Oc) -- ++(\sC-\sW:\sR) arc (\sC-\sW:\sC+\sW:\sR) -- cycle;
        \draw[hcal!85,  line width=1.15pt, opacity=0.88] (Oc) -- ++(\sC-\sW:\sR);
        \draw[hcal!85,  line width=1.15pt, opacity=0.88] (Oc) -- ++(\sC+\sW:\sR);
        \draw[hcal!85,  line width=1.15pt, opacity=0.82] (Oc) ++(\sC-\sW:\sR) arc (\sC-\sW:\sC+\sW:\sR);
        \foreach \org/\ang/\col in {Oa/\sA/ecal, Ob/\sB/track, Oc/\sC/hcal}{
          \foreach \j in {1,...,6}{
            \pgfmathsetmacro{\dang}{-6 + \j*2.2}
            \path (\org) ++(\ang:0.35) coordinate (S0);
            \path (\org) ++(\ang+\dang:1.62) coordinate (S1);
            \draw[\col!85, thin, opacity=0.45] (S0) .. controls +(0.12,0.22) and +(-0.05,0.05) .. (S1);
          }
        }
        \def\JtwoA{-38} \def\JtwoB{-12}
        \fill[qgcolor, opacity=0.14] (0,0) -- (\JtwoA:\R) arc (\JtwoA:\JtwoB:\R) -- cycle;
        \draw[qgcolor!75, line width=1.10pt, opacity=0.60] (0,0) -- (\JtwoA:\R);
        \draw[qgcolor!75, line width=1.10pt, opacity=0.60] (0,0) -- (\JtwoB:\R);
        \draw[qgcolor!75, line width=1.10pt, opacity=0.45] (\JtwoA:\R) arc (\JtwoA:\JtwoB:\R);
        \def\JtwoAxis{-24}
        \coordinate (SPLIT2) at (\JtwoAxis:0.85);
        \draw[qgcolor!70, line width=0.8pt, opacity=0.22] (0,0) -- (SPLIT2);
        \fill[qgcolor!60, opacity=0.35] (SPLIT2) circle (1.8pt);
        \coordinate (P2a) at ($(SPLIT2) + (-0.05,0.02)$);
        \coordinate (P2b) at ($(SPLIT2) + ( 0.03,0.04)$);
        \coordinate (P2c) at ($(SPLIT2) + ( 0.06,-0.02)$);
        \def\tR{1.28} \def\tW{5.2} \def\tA{-34} \def\tB{-22} \def\tC{-14}
        \foreach \org/\ang/\col in {P2a/\tA/hcal, P2b/\tB/track, P2c/\tC/ecal}{
          \fill[\col, opacity=0.16] (\org) -- ++(\ang-\tW:\tR) arc (\ang-\tW:\ang+\tW:\tR) -- cycle;
          \draw[\col!80, line width=1.00pt, opacity=0.58] (\org) -- ++(\ang-\tW:\tR);
          \draw[\col!80, line width=1.00pt, opacity=0.58] (\org) -- ++(\ang+\tW:\tR);
          \draw[\col!80, line width=1.00pt, opacity=0.52] (\org) ++(\ang-\tW:\tR) arc (\ang-\tW:\ang+\tW:\tR);
        }
        \end{scope}
        \end{tikzpicture}
    }
    \label{fig:jet_generated}
\end{subfigure}
\hfill
\begin{subfigure}[b]{0.49\linewidth}
    \centering
    \includegraphics[width=\linewidth]{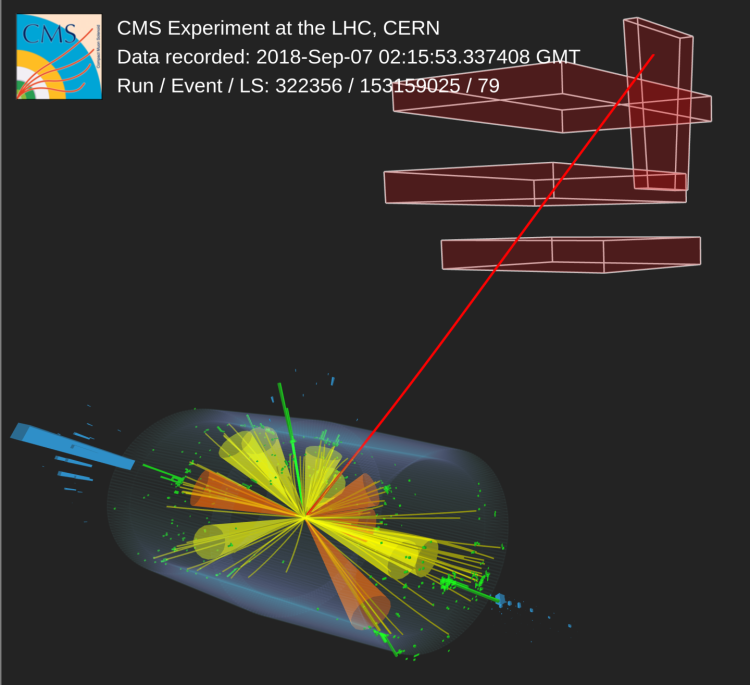}
    \caption{}
    \label{fig:jet_reference}
\end{subfigure}
\caption{\textbf{Left:} Simplified schematic of a proton-proton collision viewed transverse to the beam axis, illustrating multiple jets (colored clusters) emanating from the interaction point. \textbf{Right:} Event display at the LHC, showing reconstructed particle tracks and calorimeter energy deposits from a real collision event~\cite{CMS:2023FourTopEventDisplay}.}
\label{fig:circular_collision_comparison}
\end{figure}
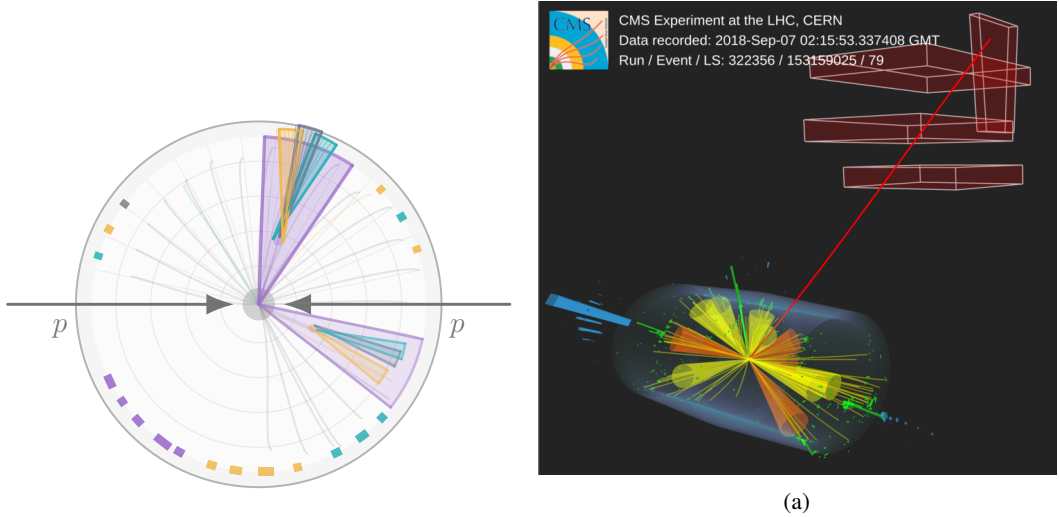

At the Large Hadron Collider (LHC), protons traveling at nearly the speed of light collide 40 million times per second, generating raw data streams in excess of 1 Petabyte per second~\cite{Evans:2008zzb,CMSP2L1T}. These collisions recreate the extreme energy conditions of the early universe, producing unstable particles that existed microseconds after the Big Bang. Hidden within this torrent of familiar physics processes are extremely rare traces of new physics, and potential signatures of dark matter. Detecting these rare signals requires distinguishing them from overwhelming backgrounds in real time, a task that requires algorithms that are both extraordinarily fast and accurate~\cite{Deiana:2021niw,Harris:2022qtm}. Figure~\ref{fig:circular_collision_comparison} illustrates this setting: a proton-proton collision at the LHC produces multiple jets emanating from a single interaction point (left), and the corresponding detector image (right) shows reconstructed tracks and calorimeter deposits that the trigger must classify within microseconds.

The challenge is formidable: high-energy particle collisions at modern accelerators produce data at rates far beyond what can be permanently stored. At the LHC, detectors generate raw data streams at the level of $\mathcal{O}(1~\mathrm{PB/s})$~\cite{Evans:2008zzb,CMSP2L1T}. Since only a small fraction of these events can be recorded, real-time data filtering is performed by trigger systems that must decide, within $\mathcal{O}(10~\mu\mathrm{s})$, whether a collision event is potentially interesting for physics analyses~\cite{CMSL1T,ATLASL1T}. The quality of these decisions directly determines which physical processes can be studied. A central task for these real-time systems is jet tagging. Jets are collimated sprays of stable particles that are produced when short-lived particles decay. Many particles produced in high-energy collisions are quarks or gluons, which cannot be observed directly due to color confinement. They instead hadronize into collimated sprays of particles that may be clustered into \emph{jets}~\cite{Cacciari:2008gp,Cacciari:2011ma}. The internal structure of a jet encodes information about the underlying quark or gluon from which the jet originates and its decay history~\cite{Larkoski:2017jix}. Accurately identifying these patterns is essential for selecting rare signals and suppressing overwhelming background processes. Even small improvements in tagging performance at the trigger level translate into large gains in recorded signal yields and physics sensitivity, directly improving discovery potential for a global community of over ten thousand scientists.

The combined data rate and latency requires ultra-fast inference, and thus models need to be both small and extremely fast to fit on dedicated FPGA hardware. Despite recent advances in jet-tagging performance, a substantial gap remains between models that achieve state-of-the-art accuracy and those that fit within this budget. Highly expressive architectures such as the Particle Transformer~\cite{Qu2022} provide excellent discriminative power, but their size and quadratic computational complexity make them impractical under these strict constraints. Conversely, compact models that satisfy the resource limits, such as small multilayer perceptrons, lack sufficient representational capacity to distinguish complex jets. Closing this gap is critical to improving particle physics data quality, and it motivates the development of physics-inspired model architectures that explicitly encode geometric and structural priors related to jet tagging, rather than scaled-down versions of large, general-purpose networks~\cite{sanh2019distilbert, howard2017mobilenets, tan2019efficientnet}. In practice, the path from a trained model to a validated FPGA deployment is extremely time- and labor-intensive. The original Transformer implementation in \texttt{hls4ml} took over two years from algorithm development to full FPGA deployment. This has driven a growing effort in the trigger ML community to first identify promising architectures at the algorithmic level, and then invest in hardware synthesis only for the most effective candidates. This is especially important where model capacity is severely limited and inductive bias has an increased effect on performance, since compact trigger models cannot rely on scale and sheer parameter count to compensate for architecture. The core challenge is therefore to design models that are both efficient by construction and sufficiently expressive for jet tagging in this resource-constrained regime.

We make four key contributions. (1) We propose PHAT-JeT, an attention-based architecture that retains \textbf{exact pairwise interactions within local patches} under tight resource constraints, complementing existing efficient transformers that either approximate attention through low-rank or cluster representations or operate at substantially higher compute. (2) We introduce a \textbf{geometric message-passing module} that encodes local detector-plane structure, improving performance across multiple attention-based architectures. (3) We demonstrate that \textbf{patch-based attention with hierarchical global communication} preserves fine-grained particle interactions at near-linear cost while remaining robust to the choice of consistent training ordering. (4) We \textbf{validate the design through ablations and four-benchmark evaluation} (\textsc{hls4ml}, JetClass, Top Tagging, Quark--Gluon) including FLOPs-matched comparisons (Appendix~\ref{app:scaled_baselines}). A glossary of jet-physics terminology in machine-learning language is provided in Appendix~\ref{app:glossary} for readers from outside the HEP community.
\section{Related Work}

\textbf{Jets as point clouds.} A jet is naturally represented as an unordered set of particles, forming a point cloud in momentum space, which connects jet tagging to the broader literature on point-cloud learning. Permutation-equivariant architectures such as PointNet~\cite{qi2017pointnetdeeplearningpoint} and deep sets~\cite{ds} aggregate per-particle features through symmetric pooling functions, and energy flow networks~\cite{efn} adapt this idea to jet physics by enforcing infrared-collinear safety in the aggregation. These set-based architectures are computationally efficient, but their reliance on global pooling limits their ability to model multi-scale correlations between particles. Graph-based methods address this limitation by learning relational features between constituents. JEDI-net~\cite{jedi} constructs interaction networks over particle pairs, and ParticleNet~\cite{Qu:2019gqs} applies dynamic graph convolutions on $k$-nearest-neighbor graphs in $(\eta,\phi)$. Both achieve strong accuracy but at a computational cost incompatible with the compute budget of real-time triggers. More recent point-cloud transformers such as Point Transformer V3~\cite{wu2024pointtransformerv3simpler} achieve scalability through spatial serialization, ordering points along a space-filling curve before applying local attention, which introduces sequential dependencies that limit parallelism on hardware.

\textbf{Transformer-based jet taggers.} Self-attention has become the dominant mechanism for capturing global correlations among jet constituents. The Particle Transformer (ParT)~\cite{Qu2022} applies full self-attention to particle clouds with pairwise interaction features and obtains the strongest reported results to date on JetClass and Top Tagging. Symmetry-aware variants build relativistic structure directly into the architecture. LorentzNet~\cite{gong2022lorentznet} constructs Lorentz-equivariant graph layers using Minkowski inner products, and the Geometric Algebra Transformer of Spinner et al.~\cite{spinner2024lorentzequivariantgeometricalgebratransformers} embeds attention in the Clifford algebra of the Lorentz group, providing exact equivariance under boosts and rotations. These models achieve excellent accuracy when compute is unconstrained, but all of them scale quadratically with the number of particles and require parameter and FLOP counts several orders of magnitude above the trigger regime, which precludes real-time deployment. We therefore treat them as \emph{full-scale baselines} that bound the achievable accuracy when resource constraints are relaxed.

\textbf{Reducing the cost of attention.} A range of efficient attention variants reduce the quadratic cost of self-attention while attempting to preserve its modeling capacity. Linformer~\cite{wang2020linformer} approximates the attention matrix with a low-rank projection along the sequence dimension, achieving linear complexity but compressing the full pairwise structure into a small number of latent components, which removes explicit particle-particle interactions. JEDI-Linear~\cite{que2025jedilinearfastefficientgraph} reformulates interaction-network message passing in linear time by replacing explicit pairwise edge MLPs with shared per-node transformations followed by global aggregation and is the strongest deployable baseline at the trigger scale. By eliminating the explicit pairwise edge computation, however, it gives up the fine-grained relational structure that motivates attention-based architectures in the first place. Physics-informed designs include SAL-T~\cite{anonymous2025salt}, a Linformer-style linear-attention transformer augmented with spatial partitioning of particles into kinematic regions and convolutional layers for local correlations, and HEPT~\cite{Miao:2024oqy,Govil:2025nvy}, which uses locality-sensitive hashing to attain near-linear complexity. Both reduce cost effectively, but SAL-T inherits the low-rank compression of its Linformer backbone, and HEPT relies on stochastic hashing whose neighborhood structure becomes unstable at the small patch sizes.

\textbf{Patch- and cluster-based attention.} Patching and clustering have been studied extensively in vision and set learning as a way to bound the cost of attention. Set Transformer~\cite{lee2019settransformer} and PaCa-ViT~\cite{grainger2023pacavit} replace dense attention with attention over a small number of learned inducing points or clusters, routing all token interactions through a low-rank summary. This reduces cost substantially but discards the exact pairwise interactions between tokens. A separate line of work, including Swin~\cite{liu2021swin} and Longformer~\cite{beltagy2020longformer}, takes a block-sparse approach by partitioning tokens into fixed-size groups and computing exact attention within each, but these methods rely on a regular spatial grid (Swin) or a meaningful sequential ordering (Longformer) to define their groups, neither of which transfers cleanly to unordered particle sets. In high-energy physics, HEP-JEPA~\cite{bardhan2025hepjepa} applies patch-level processing in a self-supervised pretraining setting where farthest-point sampling selects center particles around which context and target blocks are formed, and the context encoder predicts the latent embeddings of masked target blocks. The encoder is large by trigger standards and is intended for downstream fine-tuning rather than inference under hard latency constraints. Standard ViT patchification~\cite{dosovitskiy2020vit} similarly collapses image patches into single tokens, while sequence-preserving alternatives such as the depthwise positional encoders of CPVT~\cite{chu2021cpvt} and Point Transformer V3~\cite{wu2024pointtransformerv3simpler} inject local context without merging tokens, but operate on a regular image grid or a voxelized 3D point cloud rather than on the irregular $(\eta,\phi)$ angular plane characteristic of particle jets.
\begin{table}[h!]
\centering
\footnotesize
\setlength{\tabcolsep}{4pt}
\renewcommand{\arraystretch}{1.05}
\caption{Positioning of PHAT-JeT against representative patch- and cluster-based attention methods. ``Exact intra-region'': pairwise interactions within a region are preserved (\ding{51}) rather than routed through a low-rank or cluster bottleneck (\ding{55}).}
\begin{tabular}{lcccc}
\toprule
\textbf{Method} & \textbf{Domain} & \textbf{Approach} & \textbf{Exact Attention} & \textbf{Trigger-scale} \\
\midrule
Set Transformer~\cite{lee2019settransformer} & Sets & Inducing points (low-rank) & \ding{55} & \ding{55} \\
PaCa-ViT~\cite{grainger2023pacavit} & Vision & Patch-to-cluster (learned clusters) & \ding{55} & \ding{55} \\
Swin~\cite{liu2021swin} & Vision & Shifted local windows & \ding{51} & \ding{55} \\
HEP-JEPA~\cite{bardhan2025hepjepa} & Jets & FPS group tokens (clustered) & \ding{55} & \ding{55} \\
\textbf{PHAT-JeT (ours)} & \textbf{Jets} & \textbf{Block-sparse patch + global token} & \ding{51} & \ding{51} \\
\bottomrule
\end{tabular}
\label{tab:positioning}
\end{table}

\section{PHAT-JeT}
\label{sec:method}
\begin{figure}[!ht]
\centering
\resizebox{\columnwidth}{!}{%
\begin{tikzpicture}[
    font=\sffamily\bfseries,
    >=Stealth,
    node distance=0.5cm,
    /utils/exec={
        \definecolor{phatYellow}{RGB}{250, 229, 136}
        \definecolor{phatCyan}{RGB}{153, 235, 255}
        \definecolor{phatBlue}{RGB}{135, 206, 250}
        \definecolor{phatGreen}{RGB}{166, 226, 157}
        \definecolor{phatOrange}{RGB}{255, 179, 128}
        \definecolor{phatGrey}{RGB}{245, 245, 245}
        \definecolor{phatDarkGrey}{RGB}{80, 80, 80}
    },
    block/.style={
        draw=black, very thin, rounded corners=3pt,
        minimum height=0.7cm, minimum width=1.8cm,
        align=center, font=\footnotesize\bfseries, drop shadow
    },
    linear/.style={block, fill=phatYellow},
    attn/.style={block, fill=phatBlue, minimum height=1.2cm, minimum width=2.2cm},
    global_attn/.style={block, fill=phatOrange, minimum height=1.2cm, minimum width=2.2cm},
    patch/.style={block, fill=phatGreen, minimum width=4cm},
    container/.style={
        draw=black, fill=phatGrey, rounded corners=10pt,
        inner sep=0.3cm, drop shadow
    },
    conn/.style={->, thick, phatDarkGrey, rounded corners=2pt},
    res_conn/.style={->, thick, phatDarkGrey, dashed, rounded corners=5pt}
]
\node[container, minimum width=7.5cm, minimum height=6.5cm, fill=white] (mainBox) at (0, 2.5) {};
\node[font=\footnotesize] (inputNode) at (0, -0.3) {\textbf{Input} $X$};
\node[patch, above=0.4cm of inputNode] (patching) {Patch Partitioning};
\draw[conn] (inputNode) -- (patching);
\node[linear, above=0.5cm of patching, minimum width=3.5cm] (linProj) {Linear};
\draw[conn] (patching) -- (linProj);
\coordinate (attnCenter) at (0, 3.5);
\node[attn, anchor=east] (localAttn) at ($(attnCenter) + (-0.8, 0)$) {Local Attention\\(Intra-Patch)};
\node[global_attn, anchor=west] (globalAttn) at ($(attnCenter) + (0.8, 0)$) {Global Attention\\(Patch Tokens)};
\draw[conn] (linProj.north) -- ++(0, 0.4) -| (localAttn.270);
\draw[conn, dashed] (linProj.north) -- ++(0, 0.4) -| (globalAttn.270);
\draw[<->, thick, phatDarkGrey] (localAttn) -- node[above, font=\scriptsize\bfseries, align=center, midway] {Token\\Broadcast} (globalAttn);
\node[linear, above=1.2cm of attnCenter, minimum width=5cm] (linOut) {Linear Output};
\draw[conn] (localAttn.north) -- ++(0, 0.3) -| (linOut.200);
\draw[conn] (globalAttn.north) -- ++(0, 0.3) -| (linOut.340);
\draw[res_conn] (inputNode.west) -- ++(-3.0, 0) |- (linOut.west);
\node[font=\tiny, rotate=90, color=phatDarkGrey] at (-3.3, 2.5) {Residual Connection};
\node[below=0.2cm of mainBox.south, font=\bfseries] {PHAT-JeT Block Detail};
\begin{scope}[shift={(6.5, 0.5)}]
    \node[font=\bfseries\scriptsize] (x_in) at (0, -0.2) {Input};
    \node[block, fill=phatOrange!50, minimum width=2.5cm] (gmp_layer) at (0, 0.9) {GMP Module};
    \node[block, fill=phatYellow!50, minimum width=2.5cm, minimum height=1.2cm] (phat_block) at (0, 2.3) {PHAT Block};
    \node[block, fill=phatGreen!50, minimum width=2.5cm] (pool) at (0, 3.5) {Global Pool};
    \node[block, fill=phatCyan!50, minimum width=2.5cm] (mlp) at (0, 4.4) {MLP};
    \draw[->, thick] (mlp) -- ++(0, 0.6) node[above, font=\scriptsize] {Class Probabilities};
    \draw[conn] (x_in) -- (gmp_layer);
    \draw[conn] (gmp_layer) -- (phat_block);
    \draw[conn] (phat_block) -- (pool);
    \draw[conn] (pool) -- (mlp);
    \draw[res_conn] (x_in.east) -- ++(1.4, 0) |- (gmp_layer.east);
    \draw[res_conn] (gmp_layer.west) -- ++(-1.4, 0) |- (phat_block.west);
    \draw[gray, thin, dashed] (mainBox.north east) -- (phat_block.north west);
    \draw[gray, thin, dashed] (mainBox.south east) -- (phat_block.south west);
    \node[below=3.15cm of phat_block, font=\bfseries] {Full Pipeline};
\end{scope}
\end{tikzpicture}%
}
\caption{Architecture of PHAT-JeT. (Left) The PHAT block partitions input particles into patches before projection. (Right) The full pipeline includes geometric message passing (GMP) prior to the attention block.}
\label{fig:phat_arch_patchfirst}
\end{figure}
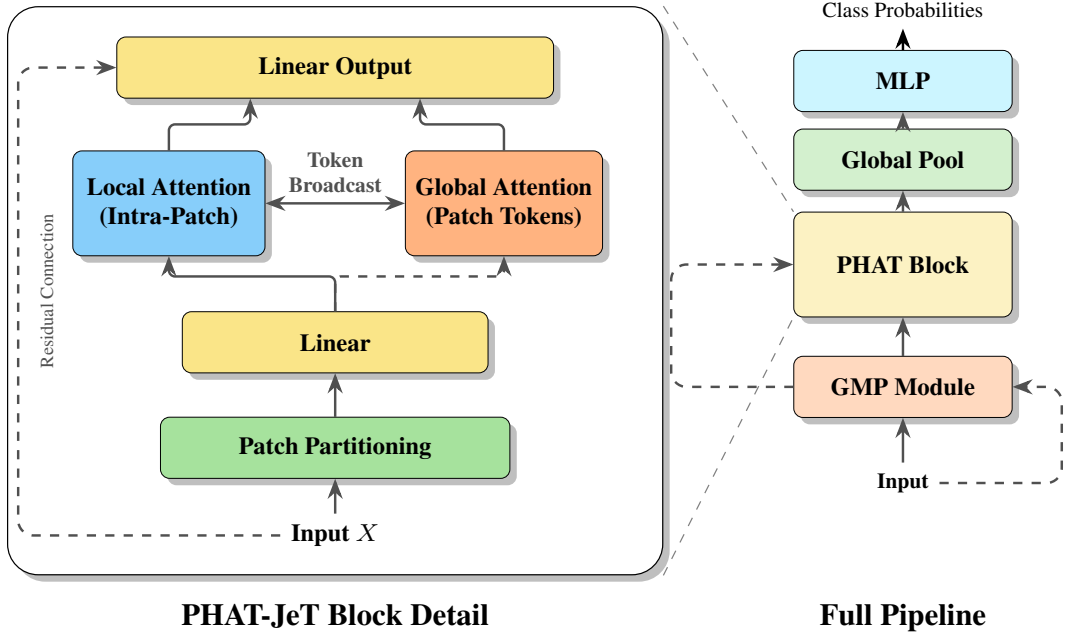

PHAT-JeT is built from three complementary components, illustrated in Figure~\ref{fig:phat_arch_patchfirst}. A \emph{Geometric Message Passing} (GMP) module injects local angular structure into particle embeddings by propagating information between neighboring regions in the $(\eta,\phi)$ plane. \emph{Local Patch-Based Self-Attention} then partitions the particle set into fixed-size patches and computes exact attention within each, reducing complexity from $\mathcal{O}(N^2)$ to $\mathcal{O}(N \cdot P)$. \emph{Hierarchical Patch-Level Attention} restores global communication by pooling each patch into a representative token, applying attention across tokens, and broadcasting the result back. Figure~\ref{fig:phat_schematic} shows how the GMP grid and the two-stage attention interact at the particle level. Empirically, restricting exact attention to patches does not sacrifice expressivity. Patch-based attention without GMP outperforms the full $\mathcal{O}(N^2)$ Transformer (81.51\% vs.\ 81.27\%, Table~\ref{tab:ablations_big_blocks_desc}) at roughly half the FLOPs.

\subsection*{Geometric Message Passing}
Decays of heavy particles of interest produce characteristic multi-prong energy deposits within a single jet cone, which appear as localized clusters in the $(\eta,\phi)$ plane, and capturing these angular correlations is essential for jet tagging. Transformer models are permutation-equivariant and therefore require an explicit notion of position to distinguish between particles; for jets, this notion is naturally geometric. We therefore introduce a learned \emph{geometric message-passing} module that injects local angular context into the particle embeddings using a lightweight 2D convolution over a coarse detector grid.

Let $X \in \mathbb{R}^{N\times C}$ denote the particle embeddings for a single jet, with coordinates $(\eta_i,\phi_i)$. We quantize each particle to a detector grid of spacing $\delta$ using per-jet minimum shifts
\begin{equation}
u_i = \left\lfloor \frac{\eta_i - \eta_{\min}}{\delta} \right\rfloor,
\qquad
v_i = \left\lfloor \frac{\phi_i - \phi_{\min}}{\delta} \right\rfloor,
\end{equation}
where $\eta_{\min}=\min_j \eta_j$ and $\phi_{\min}=\min_j \phi_j$ ensure nonnegative indices. We construct a 2D feature map by summing particle features within each grid cell
\begin{equation}
\label{eq:gmp_sum}
G_{u,v} = \sum_{i:(u_i,v_i)=(u,v)} X_i ,
\end{equation}
yielding $G \in \mathbb{R}^{H\times W\times C}$ for appropriate grid dimensions $(H,W)$. A depthwise 2D convolution is applied to the grid representation
\begin{equation}
\tilde{G} = \mathrm{DWConv}(G).
\end{equation}
We then sample the convolved grid back at the particle locations
\begin{equation}
Z_i = \tilde{G}_{u_i,v_i},
\end{equation}
and apply a pointwise channel mixing (shared across particles)
\begin{equation}
\label{eq:gmp_residual}
Y_i = W_p Z_i ,
\end{equation}
where $W_p \in \mathbb{R}^{C\times C}$ is a learned linear projection. Finally, the positional features are normalized and added residually to the input embeddings.

The GMP module can be interpreted as applying a small learnable spatial filter to a coarse representation of the jet in the $(\eta,\phi)$ plane. Each grid cell aggregates the features of particles that fall within the same angular region, forming a low-resolution map of the jet's energy and feature distribution. The convolution then propagates information between neighboring angular regions, allowing each particle to incorporate context from its surrounding area in the detector. GMP therefore encodes local geometric structure and correlations that reflect the underlying decay topology while remaining computationally inexpensive and consistent under any fixed training ordering. Subsequent patch interactions operate on geometry-aware embeddings rather than raw kinematic coordinates.

\begin{figure*}[!ht]
\centering
\hspace*{-1.7cm} % Move the whole figure left. Adjust: -0.3cm, -0.8cm, -1.0cm, etc.
\begin{tikzpicture}[
    font=\sffamily,
    >=Stealth,
    particle/.style={circle, fill=blue!60!cyan, inner sep=1.4pt, blur shadow={shadow blur steps=2}},
    enhanced_particle/.style={particle, fill=green!60!black, inner sep=1.6pt},
    gridcell/.style={draw=gray!30, fill=white, thin},
    activecell/.style={draw=orange, thick, fill=orange!16},
    patchbox/.style={rounded corners=4pt, draw=gray!20, fill=gray!5, dashed},
    token/.style={circle, draw=orange!90, fill=orange!10, thick, inner sep=2pt, minimum size=0.65cm,
                  font=\scriptsize\bfseries, blur shadow},
    section_label/.style={font=\small\bfseries, color=black!90},
    flow_arrow/.style={->, line width=1.5pt, color=gray!45},
    local_attn/.style={blue!55, line width=0.7pt, opacity=0.70},
    local_attn_arrow/.style={local_attn, <->},
    cellPink/.style={draw=magenta!80, fill=magenta!10, thick},
    cellOrange/.style={draw=orange!90, fill=orange!10, thick},
    jetBlack/.style={draw=black, thin, fill=gray!5, opacity=0.8},
    subjetPink/.style={draw=magenta!80, thick, fill=magenta!20},
    subjetOrange/.style={draw=orange!90, thick, fill=orange!20},
    gridLine/.style={gray!40, thin}
]
\begin{scope}[local bounding box=gmp, scale=1.12, transform shape]
    \node[section_label, anchor=south west] at (-0.55, 3.85) {(a) Geometric Message Passing (GMP)};
    \draw[gridLine] (0,0) grid[step=0.6] (3.0, 3.0);
    \draw[->, gray!60, thick] (0,0) -- (3.2,0) node[right, font=\tiny] {$\eta$};
    \draw[->, gray!60, thick] (0,0) -- (0,3.2) node[above, font=\tiny] {$\phi$};
    \draw[cellPink] (0.0, 1.2) rectangle (0.6, 1.8);
    \coordinate (c1) at (0.3, 1.5);
    \draw[cellOrange] (1.8, 0.0) rectangle (2.4, 0.6);
    \coordinate (c2) at (2.1, 0.3);

    \begin{scope}[shift={(c1)}]
        \def\coneL{2.1} \def\coneR{0.75} \def\subR{0.22} \def\persp{0.25}
        \draw[black, thin] (0,0) -- (\coneL, \coneR);
        \draw[black, thin] (0,0) -- (\coneL, -\coneR);
        \draw[black, thin, fill=white, fill opacity=0.5] (\coneL, 0) ellipse ({\persp} and \coneR);
        \begin{scope}[rotate=12]
            \pgfmathsetmacro{\subPersp}{(\subR/\coneR)*\persp}
            \draw[subjetPink, fill opacity=0.5] (0,0) -- (\coneL, \subR);
            \draw[subjetPink, fill opacity=0.5] (0,0) -- (\coneL, -\subR);
            \draw[subjetPink, fill=white] (\coneL, 0) ellipse ({\subPersp} and \subR);
        \end{scope}
    \end{scope}

    \begin{scope}[shift={(c2)}]
        \def\coneL{2.1} \def\coneR{0.75} \def\subR{0.22} \def\persp{0.25}
        \draw[black, thin] (0,0) -- (\coneL, \coneR);
        \draw[black, thin] (0,0) -- (\coneL, -\coneR);
        \draw[black, thin, fill=white, fill opacity=0.5] (\coneL, 0) ellipse ({\persp} and \coneR);
        \begin{scope}[rotate=-13]
            \pgfmathsetmacro{\subPersp}{(\subR/\coneR)*\persp}
            \draw[subjetOrange, fill opacity=0.5] (0,0) -- (\coneL, \subR);
            \draw[subjetOrange, fill opacity=0.5] (0,0) -- (\coneL, -\subR);
            \draw[subjetOrange, fill=white] (\coneL, 0) ellipse ({\subPersp} and \subR);
        \end{scope}
    \end{scope}

    \node[anchor=north, font=\scriptsize, align=center, text width=4.9cm, color=gray!60] at (1.5, -0.45)
        {GMP encapsulates characteristic\\ multi-prong energy flow};
\end{scope}

\draw[flow_arrow] ($(gmp.east) + (0.25, 1.25)$) -- ($(gmp.east) + (1.55, 1.25)$)
    node[midway, above, font=\tiny\bfseries, color=black!60] {DWConv + MLP};

\begin{scope}[shift={(8.0, 0)}, local bounding box=attn]
    \node[section_label, anchor=south west] at (0, 4.35) {(b) Hierarchical Attention Layers};
    \def\tokY{3.30}
    \def\patchY{0.05}
    \def\dx{3.05}
    \def\labXin{0.05}

    \node[token] (t1) at (0.0, \tokY) {$p_1$};
    \node[token] (t2) at (\dx, \tokY) {$p_2$};
    \node[token] (t3) at (2*\dx, \tokY) {$p_3$};

    \draw[<->, orange, thick] (t1) -- (t2);
    \draw[<->, orange, thick] (t2) -- (t3);
    \draw[<->, orange, thick, bend left=26] (t1) to (t3);

    \node[font=\tiny\bfseries, color=orange] at ($(t1)!0.5!(t3) + (0,0.55)$)
        {Inter-Patch (Global)};

    \node[anchor=west, font=\scriptsize\bfseries, color=blue!55]
        at (\labXin+0.25, \patchY+0.85) {Intra-Patch (Local)};

    \foreach \x/\idx in {0/1, \dx/2, {2*\dx}/3} {
        \begin{scope}[shift={(\x, \patchY)}]
            \node[patchbox, minimum width=2.35cm, minimum height=1.35cm] (pb\idx) {};

            \node[enhanced_particle] (p\idx a) at (-0.70, 0.40) {};
            \node[enhanced_particle] (p\idx b) at ( 0.55, 0.42) {};
            \node[enhanced_particle] (p\idx c) at ( 0.20,-0.40) {};
            \node[enhanced_particle] (p\idx d) at (-0.15, 0.05) {};
            \node[enhanced_particle] (p\idx e) at ( 0.05, 0.30) {};

            \draw[local_attn_arrow] (p\idx a)--(p\idx b);
            \draw[local_attn_arrow] (p\idx a)--(p\idx c);
            \draw[local_attn_arrow] (p\idx a)--(p\idx d);
            \draw[local_attn_arrow] (p\idx a)--(p\idx e);
            \draw[local_attn_arrow] (p\idx b)--(p\idx c);
            \draw[local_attn_arrow] (p\idx b)--(p\idx d);
            \draw[local_attn_arrow] (p\idx b)--(p\idx e);
            \draw[local_attn_arrow] (p\idx c)--(p\idx d);
            \draw[local_attn_arrow] (p\idx c)--(p\idx e);
            \draw[local_attn_arrow] (p\idx d)--(p\idx e);

            \node[font=\tiny\bfseries, gray!50, anchor=south]
                at ($(pb\idx.south)+(0,-0.08)$) {Patch \idx};
        \end{scope}
    }

    \foreach \tok/\pb in {t1/pb1, t2/pb2, t3/pb3} {
        \draw[->, dashed, color=gray!45, thick]
            ($(\pb.north) + (-0.30, 0.02)$) -- ($(\tok.south) + (-0.12, -0.02)$);
        \draw[->, solid, color=orange!75, line width=1.05pt]
            ($(\tok.south) + (0.12, -0.02)$) -- ($(\pb.north) + (0.30, 0.02)$);
    }

    \node[font=\tiny, color=gray!55, rotate=90]   at (-0.42, 2.05) {Mean Pool};
    \node[font=\tiny, color=orange!85, rotate=90] at ( 0.38, 2.05) {Broadcast};

    \node[anchor=north, font=\scriptsize, align=center, text width=8.6cm, color=gray!60]
        at (\dx, -0.78)
        {Complexity $O(N \cdot P)$: Local patches process independently,\\
         bridged by global interaction across patch tokens.};
\end{scope}
\end{tikzpicture}%
\caption{Particle Hierarchical Attention Transformer (PHAT-JeT) schematic. (a) GMP discretizes the $(\eta,\phi)$ plane and mixes nearby cells, capturing multi-prong substructure. (b) Exact local attention within patches and lightweight global interaction via patch tokens.}
\label{fig:phat_schematic}
\end{figure*}

\subsection*{Local Patch-Based Self-Attention}
\label{subsec:local_patch_attention}
To make attention tractable under strict resource constraints, PHAT-JeT adopts a patch-based self-attention mechanism inspired by Point Transformer V3~\cite{wu2024pointtransformerv3simpler}. Unlike the Point Transformer V3 approach, which relies on spatial serialization to define local neighborhoods, PHAT-JeT uses patches as a computational abstraction to bound the size of the attention neighborhoods and control complexity, and empirically having no change in performance as long as the training and testing sorting stays consistent. 

Given a jet with $N$ particles, we choose a patch size $P \ll N$ and partition the particle set into $N_P=\lceil N/P \rceil$ non-overlapping patches. If $N$ is not divisible by $P$, we pad the input so that all patches contain exactly $P$ particles. In our formulation, patches are not intended to represent spatial regions or neighborhoods. Instead, the patch size acts as a tunable hyperparameter that directly controls the trade-off between computational cost and modeling capacity.

Within each patch, we apply a standard multi-head self-attention operation independently. Each particle attends only to the particles within the same patch, and no attention is computed across patches at this stage. In a single layer, each particle therefore interacts with only a bounded subset of the full particle set. The resulting loss of global context is addressed by the hierarchical communication stage described in the following section. Let $X^{(j)} \in \mathbb{R}^{P \times d_{\text{model}}}$ denote the features in patch $j$. We apply scaled dot-product attention within each patch:
\begin{equation}
Z^{(j)} = \mathrm{Attention}(X^{(j)} W_Q, X^{(j)} W_K, X^{(j)} W_V).
\end{equation}
Outputs from all heads are concatenated and projected with $W_O$ to produce the updated patch features. All patches are processed in parallel. Since attention is computed only within patches of size $P$, the overall complexity scales as
\begin{equation}
\mathcal{O}(N_P \cdot P^2)
=
\mathcal{O}\!\left(\left\lceil \frac{N}{P} \right\rceil P^2\right)
\;\sim\;
\mathcal{O}(N \cdot P),
\end{equation}
which is near linear in $N$ for fixed $P$.

Other approaches can reduce the cost of self-attention but differ in how interaction structure is imposed. Full global attention is computationally prohibitive at realistic particle multiplicities, while low-rank approximations such as Linformer~\cite{wang2020linformer} reduce complexity by compressing the attention matrix, which can suppress explicit particle--particle interactions. Sliding-window and serialized attention mechanisms can also be viewed as a form of patching, but rely on an assumed meaningful particle ordering to define interaction neighborhoods~\cite{anonymous2025salt}. This introduces an ordering-dependent inductive bias and sequential preprocessing that complicates parallel hardware deployment. In contrast, we show in Section~\ref{sec:results} that PHAT-JeT performs equivalently across orderings including random ordering of inputs, as long as the training and testing data are sorted the same way.(Table ~\ref{tab:order_mismatch}). Since the inputs are always sorted by $\pT$ in the detector, PHAT-JeT removes the need for additional real-time sorting. 

\subsection*{Hierarchical Patch-Level Attention}
The same patch partition used for local particle-level attention is reused to enable efficient global communication. While local patch attention allows particles within each patch to exchange information, it does not provide a mechanism for interaction between different patches. To address this, PHAT-JeT introduces a hierarchical patch-level attention stage that operates on a pooled representation of each patch.

Let the particle sequence after local attention be partitioned into $N_P=\lceil N/P \rceil$ non-overlapping patches of size $P$, using the same padding scheme as in the local attention stage. Denote the features of patch $j$ by $X^{(j)} \in \mathbb{R}^{P \times d_{\text{model}}}$. We construct a patch token $p_j \in \mathbb{R}^{d_{\text{model}}}$ by mean pooling over the particles in the patch
\begin{equation}
p_j \;=\; \frac{1}{P}\sum_{i=1}^{P} x^{(j)}_{i},
\end{equation}
where $x^{(j)}_{i} \in \mathbb{R}^{d_{\text{model}}}$ is the feature vector of the $i$-th particle in patch $j$. This patch token provides a compact summary of the information learned by the local attention stage.

We then apply multi-head self-attention to the sequence of patch tokens $P=\{p_1,\dots,p_{N_P}\}$ to obtain globally updated patch representations
\begin{equation}
\tilde{p}_j \;=\; \mathrm{MHSA}(P)_j.
\end{equation}
Since $N_P = N/P$, this global communication stage operates on a much shorter sequence than particle-level attention and has complexity $\mathcal{O}(N_P^2 d_{\text{model}}) = \mathcal{O}\!\left((N/P)^2 d_{\text{model}}\right)$, which is negligible relative to the local patch attention $\mathcal{O}(N P)$ for $P \ll N$.

Finally, each updated patch token is broadcast back to all particles in the corresponding patch. We optionally apply a learned linear projection $W_m \in \mathbb{R}^{d_{\text{model}} \times d_{\text{model}}}$ before broadcasting. For each particle $i$ in patch $j$, the patch message is
\begin{equation}
m^{(j)}_{i} \;=\; W_m \tilde{p}_j,
\end{equation}
and this message is added residually to the particle features. This hierarchical communication mechanism provides an efficient path for global information flow, while preserving the fine-grained interactions learned by the local patch attention and maintaining near-linear computational scaling.

\begin{table*}[!ht]
\centering
\small
\caption{Quantitative comparison on the \textsc{hls4ml} jet-tagging benchmark; \textbf{Avg bkg Rej} denotes the average background rejection at 80\% signal efficiency (higher is better). All ROC AUC uncertainties are below 0.001. All resource-constrained baselines use a single attention (or equivalent) layer with similar hidden dimension, with FLOPs scaled to match PHAT-JeT. Unconstrained reference (ParT~\cite{Qu2022}) is trained on the same data and operates at $\mathcal{O}(10^2)\times$ more FLOPs.}
\begin{tabular*}{\textwidth}{@{\extracolsep{\fill}}lccccc}
\toprule
\textbf{Model} & \textbf{Acc (\%)} $\uparrow$  & \textbf{ROC AUC} $\uparrow$  & \textbf{Avg bkg Rej} $\uparrow$  & \textbf{\# Params (K)} & \textbf{FLOPs (M)} \\
\midrule
\textbf{PHAT-JeT} & \textbf{81.80 $\pm$ 0.02} & \textbf{0.962} & \textbf{71.6 $\pm$ 0.6} & 6.7 & 1.31 \\
\midrule
JEDI-Linear  & 81.56 $\pm$ 0.05 & 0.961 & 68.7 $\pm$ 2.7 & 19.8 & 1.38 \\
Transformer & 81.27 $\pm$ 0.06 & 0.959 & 66.9 $\pm$ 0.6 & 4.6 & 2.48 \\
SAL-T & 81.28 $\pm$ 0.10 & 0.960 & 64.7 $\pm$ 0.5 & 5.1 & 1.26 \\
Linformer & 81.22 $\pm$ 0.07 & 0.960 & 65.0 $\pm$ 1.7 & 11.9 & 1.32 \\
HEPT~\cite{Miao:2024oqy,Govil:2025nvy} & 74.10 $\pm$ 0.02 & 0.929 & 25.7 $\pm$ 0.3 & 8.5 & 1.17 \\
PointTransformer V3 & 80.93 $\pm$ 0.10 & 0.958 & 61.6 $\pm$ 0.6 & 22.4 & 1.24 \\
PointNet & 75.04 $\pm$ 0.06 & 0.934 & 17.0 $\pm$ 0.1 & 9.8 & 1.35 \\
\midrule
\multicolumn{6}{l}{\textit{Unconstrained reference}} \\
ParT (full)~\cite{Qu2022} & 82.10 $\pm$ 0.51 & 0.964 & 72.7 $\pm$ 1.0 & 2{,}127 & 313 \\
\bottomrule
\end{tabular*}
\label{tab:performance}
\end{table*}

\section{Results}
\label{sec:results}
We evaluate PHAT-JeT on multiple jet tagging tasks and compare it against baseline models that are feasible at the trigger scale. In all experiments, transformer models are constrained to a single Transformer (or equivalent) layer with comparable model dimension, to ensure a fair comparison under tight resource budgets. We use the default JEDI-Linear setup.

\textbf{Metrics.} We report standard classification metrics such as accuracy and ROC AUC, but place particular emphasis on \textit{background rejection} at fixed signal efficiency, as this is the primary trigger-quality metric. Background rejection is defined as $1/\text{FPR}@\text{TPR}{=}0.8$ (the inverse false positive rate at a given true positive rate); higher values indicate that more background events can be rejected at the same signal acceptance rate.

\textbf{Dataset and Setup.} Our primary benchmark is the public \textsc{hls4ml} LHC Jet Dataset~\cite{pierini_2020_3602260}, which contains simulated proton--proton collision events designed to emulate the constraints of the Compact Muon Solenoid (CMS) Level-1 trigger. The dataset includes 504,000 jets for training, 126,000 for validation, and 240,000 for testing. Jets belong to five classes corresponding to their origin: light quark ($q$), gluon ($g$), $W$ boson, $Z$ boson, and top quark ($t$). Each jet is represented by the four-momentum vectors of up to 150 constituent particles. Preprocessing details are given in Appendix~\ref{app:dataset}. All models are trained using the same pipeline: categorical cross-entropy loss, the Adam optimizer, and a batch-size scheduler. We use the official training and validation splits provided with the dataset, and report performance on the held-out test set. Additional details of the training procedure are provided in Appendix~\ref{app:training}.

\textbf{Baseline models.} We compare PHAT-JeT against six representative architectures for particle-cloud or hardware-constrained inference: JEDI-Linear~\cite{que2025jedilinearfastefficientgraph}, a single-layer Linformer~\cite{wang2020linformer}, a single-layer standard Transformer~\cite{NIPS2017_3f5ee243}, SAL-T~\cite{anonymous2025salt}, Point Transformer V3~\cite{wu2024pointtransformerv3simpler}, and HEPT~\cite{Miao:2024oqy,Govil:2025nvy}. To ensure a fair comparison, all models use the same input representation, the same $\kT$-based particle ordering for non-equivariant models~\cite{Ellis:1993tq} (see Appendix~\ref{app:baselines}), four attention heads, and similar hidden dimensions across Transformer-based models. We use a patch size of $P{=}10$, resulting in 15 patches and roughly 10\% of the attention computations of a full Transformer.

\textbf{Results and Analysis.} Table~\ref{tab:performance} shows that PHAT-JeT achieves the highest accuracy, ROC AUC, and background rejection on \textsc{hls4ml} under similar FLOPs with the baseline models. The improvement over JEDI-Linear, the strongest prior FPGA-deployable baseline, is statistically significant in accuracy and background rejection. The same pattern holds on more challenging benchmarks. On JetClass, a much harder 10 class jet tagging benchmark, PHAT-JeT performs much better than any other model at this compute range (Table~\ref{tab:performance_jetclass}).  See Appendix~\ref{app:ablations} for the full hyperparameter sweep.

\begin{table*}[!ht]
\centering
\small
\caption{JetClass benchmark~\cite{Qu2022}, 10-class jet tagging. Resource-constrained models are scaled to $\sim$1.2--1.5\,M FLOPs with 128 particles per jet and trained on a 2\,M-jet subset of the 100\,M-jet training set under the available compute budget, and evaluated on the standard 20\,M-jet test set. ParT and L-GATr are included as full-scale architecture references trained on the 2\,M-jet subset as reported in their original papers. Unconstrained references operate at $\mathcal{O}(10^2$--$10^4)\times$ more FLOPs than the resource-constrained block.}
\begin{tabular*}{\textwidth}{@{\extracolsep{\fill}}lccccc}
\toprule
\textbf{Model} & \textbf{Acc (\%)} & \textbf{ROC AUC} & \textbf{Avg bkg Rej} & \textbf{\# Params (K)} & \textbf{FLOPs (M)} \\
\midrule
\textbf{PHAT-JeT} & \textbf{65.38 $\pm$ 0.08} & \textbf{0.9391 $\pm$ 0.0003} & \textbf{43.94 $\pm$ 0.65} & 6.8 & 1.24 \\
\midrule
% ptv3 & 64.69 $\pm$ 0.11 & 0.9372 $\pm$ 0.0003 & 42.21 $\pm$ 0.25 & 23.7 & 1.45 \\
Linformer & 62.64 $\pm$ 0.19 & 0.9305 $\pm$ 0.0006 & 33.24 $\pm$ 0.11 & 9.5 & 1.49 \\
SAL-T & 62.19 $\pm$ 0.28 & 0.9291 $\pm$ 0.001 & 31.78 $\pm$ 0.90 & 7.0 & 1.42 \\
JEDI-Linear & 59.52 $\pm$ 0.82 & 0.9192 $\pm$ 0.003 & 21.93 $\pm$ 1.64 & 6.1 & 1.31 \\
Small ParT & 57.58 $\pm$ 0.28 & 0.9115 $\pm$ 0.002 & 15.83 $\pm$ 0.43 & 2.9 & 1.52 \\
Transformer & 56.18 $\pm$ 1.34 & 0.9056 $\pm$ 0.006 & 15.09 $\pm$ 2.53 & 1.0 & 1.24 \\
PointNet & 55.87 $\pm$ 0.16 & 0.9063 $\pm$ 0.001 & 20.83 $\pm$ 0.70 & 9.9 & 1.35 \\
\midrule
\multicolumn{6}{l}{\textit{Unconstrained references}} \\
\midrule
ParT (full)~\cite{Qu2022} & 83.6 & 0.9834 & 2345.6 & 2{,}140 & 340 \\
L-GATr~\cite{spinner2024lorentzequivariantgeometricalgebratransformers} & 83.9 & 0.9842 & 2844.2 & 1{,}682 & 9{,}709 \\
\bottomrule
\end{tabular*}
\label{tab:performance_jetclass}
\end{table*}
 
 On the Top Tagging and Quark--Gluon benchmarks(Table~\ref{tab:performance_top_qg}), PHAT-JeT remains the strongest model. Taken together, these results indicate that the inductive biases introduced by GMP and patch-token attention generalize beyond \textsc{hls4ml}, and that PHAT-JeT with GMP is the best performing model at this compute cost across benchmarks.

\begin{table*}[!ht]
\centering
\small
\caption{Performance on the Top Tagging~\cite{Kasieczka2019TopQuark} and Quark--Gluon~\cite{komiske_2019_3164691} datasets. $^{\dagger}$Bkg Rej for resource-constrained models is $1/\mathrm{FPR}@0.8\,\mathrm{TPR}$. The full-scale references operate at $\mathcal{O}(10^2$--$10^3)\times$ more FLOPs. Dataset details in Appendix~\ref{app:top_qg}.}
\begin{tabular}{lccc|ccc}
\toprule
& \multicolumn{3}{c|}{\textbf{Top Tagging}} & \multicolumn{3}{c}{\textbf{Quark--Gluon (QG)}} \\
\textbf{Model} & \textbf{Test Acc (\%)} & \textbf{ROC AUC} & \textbf{Bkg Rej$^{\dagger}$} & \textbf{Test Acc (\%)} & \textbf{ROC AUC} & \textbf{Bkg Rej$^{\dagger}$} \\
\midrule
\multicolumn{7}{l}{\textit{Resource-constrained models (single attention layer, $\sim$1\,M FLOPs)}} \\
\midrule
\textbf{PHAT-JeT}   &  $\mathbf{92.69 \pm 0.20}$ & $\mathbf{0.979}$ & $\mathbf{34.3 \pm 2.1}$  & $\mathbf{81.80 \pm 0.09}$ & $\mathbf{0.893}$ & $\mathbf{6.1 \pm 0.1}$ \\
SAL-T               & $92.52 \pm 0.11$ & $0.978$ & $31.8 \pm 1.1$ & $81.34 \pm 0.05$ & $0.889$ & $5.8 \pm 0.1$ \\
Linformer           & $92.40 \pm 0.07$ & $0.977$ & $30.9 \pm 1.1$ & $81.36 \pm 0.01$ & $0.888$ & $5.8 \pm 0.0$ \\
JEDI-Linear         & $91.18 \pm 0.65$ & $0.971$ & $22.8 \pm 3.2$ & $80.86 \pm 0.06$ & $0.885$ & $5.5 \pm 0.1$ \\
\midrule
\multicolumn{7}{l}{\textit{Unconstrained references (full architectures, $\mathcal{O}(10^2$--$10^3)\,\mathrm{M}$ FLOPs; numbers from original papers)}} \\
\midrule
ParT~\cite{Qu2022}                         & $94.0$ & $0.986$ & $413\pm16^{\ast}$  & 84.9 & 0.920 & $47.9\pm0.5$ \\
LorentzNet~\cite{gong2022lorentznet}       & $94.2$ & $0.987$ & $498\pm18^{\ast}$  & $84.4$ & $0.916$ & $42.4\pm0.4^{\ast}$ \\
L-GATr~\cite{spinner2024lorentzequivariantgeometricalgebratransformers} & $94.2$ & $0.987$ & $548\pm26^{\ast}$  & --- & --- & --- \\
\bottomrule
\end{tabular}
\label{tab:performance_top_qg}
\end{table*}

\section{Ablation Studies}
\label{sec:ablations}
Ablation studies (Appendix~\ref{app:ablations}, Table~\ref{tab:ablations_big_blocks_desc}) validate the two core components of our design and show that GMP transfers across attention architectures. Patch-size sweeps, pooling-mode comparisons, per-class breakdowns, train/test ordering mismatches, and patch-assignment robustness are deferred to Appendices~\ref{app:ablations}.

\paragraph{Impact of Geometric Message Passing.} Adding GMP consistently improves performance across all PHAT-JeT variants. The model without GMP remains competitive in accuracy and ROC AUC but exhibits noticeably weaker background suppression, indicating that attention and ordering alone do not capture the local geometric structure of jets in the $(\eta,\phi)$ plane. Performance is stable across GMP grid spacings $\delta=0.1$--$0.2$ and degrades modestly only at $\delta=0.3$ (Appendix~\ref{app:grid_spacing}).

\paragraph{Impact of Hierarchical Patch-Level Attention.} Removing the global patch-token stage from PHAT-JeT induces a drop in average background rejection by 1--3 points across variants (Appendix~\ref{app:ablations}, Table~\ref{tab:ablations_big_blocks_desc}), confirming that local patch attention alone is insufficient and the global patch-token mechanism is essential for aggregating information across patches.

\paragraph{Robustness to Particle Sorting and Patch Assignment.} PHAT-JeT is robust to the \emph{choice} of consistent training ordering ($\kT$, $\pT$, Morton, and random orderings yield indistinguishable performance when sorting is fixed at training and test time, as shown in table \ref{tab:order_mismatch}), but train/test mismatches degrade performance, so this is not permutation invariance at deployment. Since jets are always sorted by $\pT$ in the detector, PHAT-JeT avoids any further need to further sort in real time, which was a bottleneck for previous trigger-oriented models. For a heuristic model of why PHAT-JeT may be more robust to the ordering, refer to Appendix \ref{app:coverage-model}.

\begin{table}[h!]
\centering
\small
\caption{PHAT-JeT test accuracy (\%) on \textsc{hls4ml} under train/test ordering mismatch. The row represents test set sorting, the columns represent train set sorting. Diagonal: matched train/test ordering. Off-diagonal: trained with one ordering and tested with another. The diagonal entries have indistinguishable performance across schemes (the model adapts to whichever ordering it sees during training); the off-diagonal degrades, showing that the model is not deployment-time permutation-invariant, but performance is robust if the sorting is consistent.}
\setlength{\tabcolsep}{4pt}
\begin{tabular}{lcccc}
\toprule
\textbf{Test} $\backslash$ \textbf{Train} & $\kT$ & $\pT$ & Morton & Random \\
\midrule
$\kT$    & \textbf{81.80 $\pm$ 0.04} & 74.35 $\pm$ 2.30 & 77.69 $\pm$ 3.59 & 52.74 $\pm$ 19.23 \\
$\pT$    & 79.33 $\pm$ 0.52 & \textbf{81.83 $\pm$ 0.08} & 78.28 $\pm$ 2.65 & 49.26 $\pm$ 21.33 \\
Morton   & 69.34 $\pm$ 6.69 & 55.68 $\pm$ 2.55 & \textbf{81.85 $\pm$ 0.13} & 61.88 $\pm$ 17.11 \\
Random   & 35.91 $\pm$ 10.09 & 28.70 $\pm$ 6.40 & 64.87 $\pm$ 16.42 & \textbf{81.73 $\pm$ 0.08} \\
\bottomrule
\end{tabular}
\label{tab:order_mismatch}
\end{table}

\begin{table}[h]
  \centering
  \vspace{-8pt}
  \caption{Ablation: contribution of hierarchical global communication and geometric message passing.}
  \label{tab:ablation_local_global_gmp}
  \small
  \begin{tabular}{lcc@{\hspace{3pt}}l}
    \toprule
    \textbf{Configuration} & \textbf{Acc (\%)} & \multicolumn{2}{c}{\textbf{Avg bkg Rej}} \\
    \midrule
    Local Patch only       & $81.23 \pm 0.03$ & $64.8 \pm 0.7$ & \\
    + Global patch tokens  & $81.51 \pm 0.11$ & $67.0 \pm 1.5$ & \gain{3.4\%} \\
    + GMP                  & $81.80 \pm 0.02$ & $71.6 \pm 0.6$ & \gain{6.9\%} \\
    \bottomrule
  \end{tabular}
  \vspace{-8pt}
\end{table}

\section{Conclusion}
\label{sec:conclusion}
PHAT-JeT is a hierarchical attention transformer that achieves state-of-the-art jet tagging across \textsc{hls4ml}, JetClass, Top Tagging, and Quark--Gluon at a strict trigger budget, by combining local patch-based attention with a lightweight global stage and a geometric message-passing module encoding $(\eta,\phi)$ structure, at the same scale as JEDI-Linear baseline~\cite{Duarte:2018ite}. The results show that explicit architectural priors are especially valuable in the low-capacity regime: exact intra-patch attention preserves fine-grained particle interactions, patch-token communication restores global context, and GMP injects detector-plane locality without requiring expensive graph construction or full quadratic attention. Together, these components narrow the gap between trigger-feasible models and unconstrained jet taggers while remaining simple enough to motivate future hardware synthesis studies.

\section{Acknowledgement}
A.G., and J.N. are supported by the DOE Office of Science, Award No. DE-SC0023524, FermiForward Discovery Group, LLC under Contract No. 89243024CSC000002 with the U.S. Department of Energy, Office of Science, Office of High Energy Physics, LDRD L2024-066-1, Fermilab, DOE Office of Science, Office of High Energy Physics ``Designing efficient edge AI with physics phenomena'' Project (DE-FOA-0002705), DOE Office of Science, Office of Advanced Scientific Computing Research under the ``Real-time Data Reduction Codesign at the Extreme Edge for Science'' Project (DE-FOA-0002501).

Z.Z. and J.D. are supported by the Research Corporation for Science Advancement (RCSA) under grant No. CS-CSA-2023-109, U.S. Department of Energy (DOE), Office of Science, Office of High Energy Physics under Grant No. DE-SC0009919, and the U.S. National Science Foundation (NSF) Harnessing the Data Revolution (HDR) Institute for Accelerating AI Algorithms for Data Driven Discovery (A3D3) under Cooperative Agreement No. PHY-2117997.

A.W. is supported by the Visiting Scholars Award Program of the Universities Research Association.

\newpage
\bibliography{bibliography}
\bibliographystyle{cms_unsrt}

\newpage
\appendix
\section{Glossary: jet physics terminology}
\label{app:glossary}

This appendix defines the high-energy physics (HEP) terms used in the paper in language familiar to a general machine learning audience.

\paragraph{Data representation.}
A \emph{jet} is an unordered set of 20--150 particles produced when a high-energy quark or gluon fragments; equivalently, a sparse point cloud in cylindrical coordinates $(\eta,\phi)$. Each \emph{constituent} (or particle) is a feature vector that plays the role of a token, with core features given by the four-momentum components $(\pT,\eta,\phi,m)$ and optional discrete (particle ID) and continuous (impact-parameter) attributes. The \emph{transverse momentum} $\pT$ is the momentum component perpendicular to the beam axis, which acts as an importance weight (high-$\pT$ constituents carry more discriminative information). The \emph{pseudorapidity} $\eta=-\ln\tan(\theta/2)$ and the \emph{azimuthal angle} $\phi$ together form the 2D spatial coordinate system of the point cloud, analogous to $(x,y)$ in an image.

\paragraph{Jet clustering and ordering.}
The \emph{anti-$\kT$ algorithm} is the standard jet-clustering algorithm: a hierarchical agglomerative procedure that iteratively merges particles starting from the highest-$\pT$ seed within a cone of radius $R$, producing approximately circular clusters in $(\eta,\phi)$. The \emph{$\kT$ distance} is the physics-motivated pairwise metric $d_{ij}=\min(p_{\mathrm{T},i}^{2}, p_{\mathrm{T},j}^{2})\cdot\Delta R_{ij}^{2}/R^{2}$, which weights spatial proximity by momentum and reflects the branching structure of the underlying parton shower. We use this distance to define a $\kT$-based particle ordering for our non-permutation-equivariant baselines (Section~\ref{sec:results}).

\paragraph{Jet substructure: prongs and resonances.}
The internal structure of a jet is characterized by its \emph{prong count}: QCD background jets are 1-prong, hadronic decays of $W$ or $Z$ bosons give 2-prong jets at masses $\sim80$--$91$\,GeV, Higgs decays $H{\to}b\bar{b}/c\bar{c}/gg$ give 2-prong jets at $\sim125$\,GeV, and top decays $t{\to}bW{\to}bq\bar{q}$ give 3-prong jets at $\sim173$\,GeV. Resolving prong count and resonance mass is therefore the core task of jet tagging.

\paragraph{Classification tasks.}
\emph{Jet tagging} on the \textsc{hls4ml} dataset is a 5-class problem (light quark, gluon, $W$, $Z$, top). \emph{JetClass}~\cite{Qu2022} extends this to 10 classes (QCD plus nine signal processes from top, $W$, $Z$, and Higgs decays). \emph{Top tagging}~\cite{Kasieczka2019TopQuark} is binary (top vs.\ QCD) and \emph{Quark--Gluon tagging}~\cite{komiske_2019_3164691} is binary (quark- vs.\ gluon-initiated), both serving as out-of-distribution transfer benchmarks. The primary trigger-relevant evaluation metric is \emph{background rejection at fixed signal efficiency}, $1/\varepsilon_{B}$ at a chosen $\varepsilon_{S}$ (we use $\varepsilon_{S}=0.8$ throughout the resource-constrained main results); higher is better, and this metric is more sensitive than accuracy because background dominates real data by orders of magnitude.

\paragraph{Hardware-specific terms.}
The \emph{Level-1 (L1) trigger} is the first stage of online event filtering at the LHC, implemented in custom hardware with sub-microsecond latency budgets and constrained resources (logic cells, BRAM, DSPs). \emph{FPGAs} (Field-Programmable Gate Arrays) are the standard hardware target for these trigger applications, and \emph{hls4ml}~\cite{Duarte:2018ite} is the toolchain that converts trained ML models into FPGA firmware via high-level synthesis. The \textsc{hls4ml} dataset shares its name with this toolchain because it was introduced as a representative benchmark for trigger-feasible jet tagging.

\section{Dataset}
\label{app:dataset}
We follow the preprocessing and experimental setup used in prior trigger-focused studies. Initial-state particles are generated with a transverse momentum of at least 1~TeV, and the final-state energies are smeared to approximate the CMS detector response. For each jet, we select the $n$ highest-$\pT$ particles with $\pT > 1$\,GeV. If a jet contains fewer than $n$ constituents, it is padded with zeros. As input features, we use the transverse momentum $\pT$, pseudorapidity difference $\Delta\eta$, and azimuthal angle difference $\Delta\phi$ of each particle relative to the jet axis. The $\pT$ values are normalized to the 5th and 95th percentile range computed over the training set.

\begin{figure}[h!]
    \centering
    \includegraphics[width=0.8\linewidth]{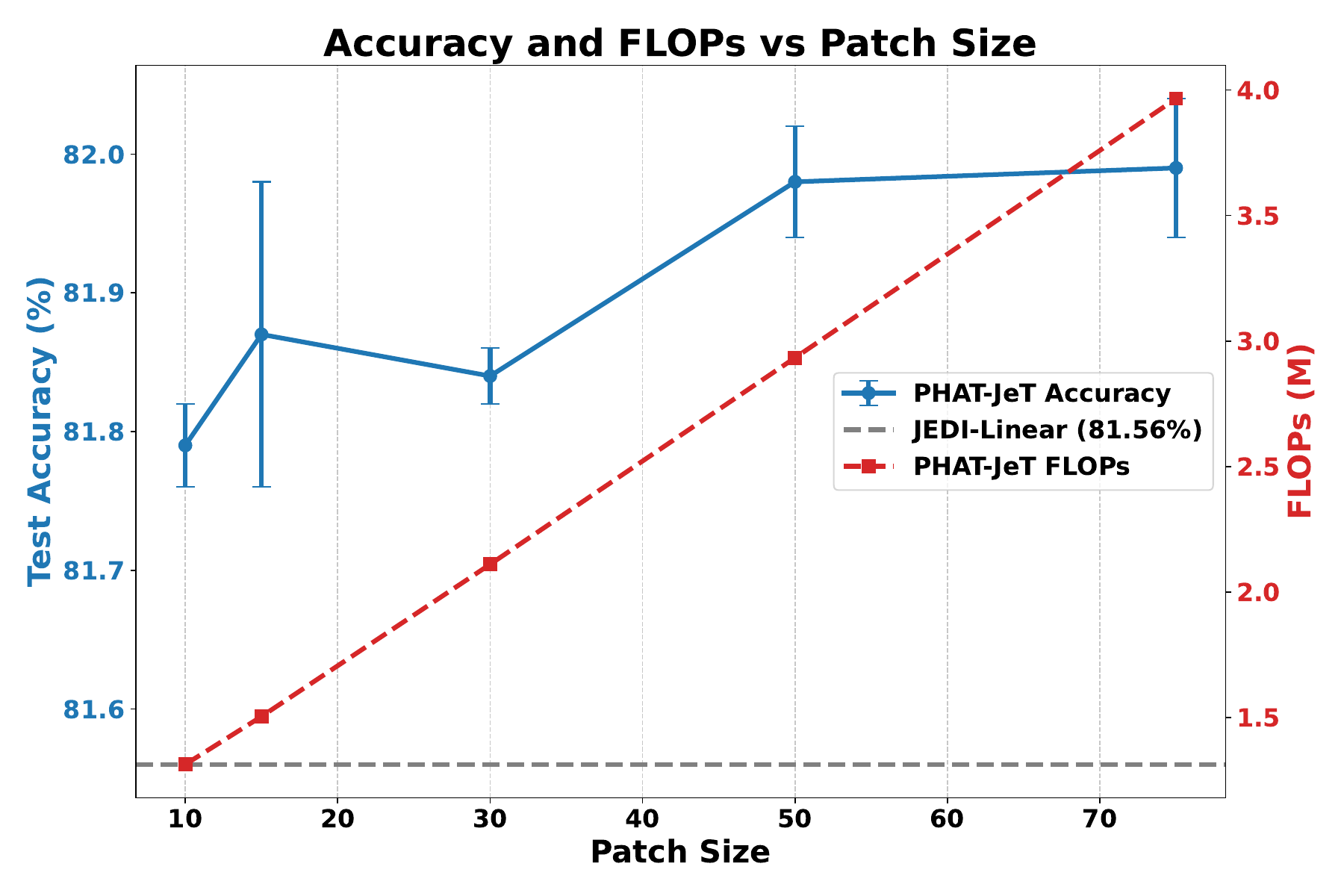}
    \caption{Accuracy and FLOPs as patch size increases for PHAT-JeT. JEDI-Linear is included as a baseline comparison.}
    \label{fig:patch_size_trend}
\end{figure}

\section{Hyperparameter scan results and ablations}
\label{app:ablations}
This appendix reports the consolidated ablation summary (Table~\ref{tab:ablations_big_blocks_desc}) and the full hyperparameter sweep that accompanies it.

\begin{table}[h!]
\centering
\footnotesize
\setlength{\tabcolsep}{4pt}
\renewcommand{\arraystretch}{0.95}
\caption{Consolidated ablation results on the \textsc{hls4ml} dataset (ROC AUC uncertainties $<0.001$). The hierarchical and GMP blocks validate PHAT-JeT's two core components and demonstrate that GMP transfers across attention architectures. The sorting summary row condenses four runs ($\kT$, $\pT$, Morton, random) into a single mean. Per-class background rejection breakdowns and the without-GMP per-architecture controls are in Tables~\ref{tab:ppt_sweep_top10_no_agg_GMP_auc3_bgrej1} and~\ref{tab:ppt_ablation_no_GMP_top10_auc3_bgrej1_noagg}; pooling-mode and train/test mismatch are in Appendix~\ref{app:pooling_mode} and Table~\ref{tab:order_mismatch}. All rows share PHAT-JeT's hidden dimension, so FLOPs differ from those in Table~\ref{tab:performance}.}
\begin{tabular}{lccccr}
\toprule
\textbf{Description} & \textbf{Acc (\%)} & \textbf{ROC AUC} & \textbf{Bg Rej} & \textbf{\# Params} & \textbf{FLOPs} \\
\midrule
\multicolumn{6}{l}{\textit{Hierarchical Patch-Level Attention (Global stage)}} \\
\textbf{PHAT-JeT (with Global)} & \textbf{81.80 $\pm$ 0.02} & \textbf{0.962} & \textbf{71.6 $\pm$ 0.6} & 6,694 & 1,313,926 \\
PHAT-JeT (without Global) & 81.76 $\pm$ 0.02 & 0.961 & 69.7 $\pm$ 1.2 & 5,045 & 924,877 \\
\midrule
\multicolumn{6}{l}{\textit{GMP transfers across attention mechanisms}} \\
\textbf{PHAT-JeT (with GMP)} & \textbf{81.80 $\pm$ 0.02} & \textbf{0.962} & \textbf{71.6 $\pm$ 0.6} & 6,694 & 1,313,926 \\
PHAT-JeT (without GMP) & 81.51 $\pm$ 0.11 & 0.960 & 67.0 $\pm$ 1.5 & 5,045 & 1,307,814 \\
PHAT-JeT (Sliding Window) + GMP & 81.73 $\pm$ 0.05 & 0.961 & 70.4 $\pm$ 1.0 & 6,405 & 1,164,596 \\
SAL-T + GMP & 81.49 $\pm$ 0.06 & 0.961 & 65.9 $\pm$ 1.5 & 4,583 & 751,770 \\
Linformer + GMP & 81.45 $\pm$ 0.02 & 0.961 & 65.0 $\pm$ 0.8 & 6,953 & 643,770 \\
\midrule
\multicolumn{6}{l}{\textit{Particle Sorting (consistent train/test ordering)}} \\
PHAT-JeT (mean over $\kT$, $\pT$, Morton, random)$^{\dagger}$ & 81.83 $\pm$ 0.04 & 0.962 & 71.8 $\pm$ 0.8 & 6,694 & 929,926 \\
\bottomrule
\end{tabular}
\\[1pt]
{\scriptsize $^{\dagger}$Per-scheme accuracies indistinguishable under consistent train/test ordering: $\kT$ $81.80\pm0.02$, $\pT$ $81.80\pm0.15$, Morton $81.86\pm0.02$, random $81.87\pm0.06$.}
\label{tab:ablations_big_blocks_desc}
\end{table}

We sweep over five binary hyperparameters that determine the patch-token communication path: \texttt{use\_pool} (whether the global pool is included), \texttt{ffn\_activation} (\texttt{gelu} or \texttt{relu}), \texttt{patch\_tokenizer\_mode} (\texttt{mean}, \texttt{learned\_pool}, \texttt{flatten\_dense}, or \texttt{max}), \texttt{message\_proj} (whether the broadcast back to particles passes through a learned linear projection $W_m$), and \texttt{message\_gated} (whether a learned scalar gate weights the message before the residual add). For each setting we run $n{=}3$ seeds and report mean$\pm$std for accuracy and per-class background rejection ($W$, $Z$, top); the average background rejection used in the main text is the mean over the three signal classes. Tables~\ref{tab:ppt_sweep_top10_no_agg_GMP_auc3_bgrej1} and~\ref{tab:ppt_ablation_no_GMP_top10_auc3_bgrej1_noagg} show the top-10 configurations under \texttt{use\_GMP=True} and \texttt{use\_GMP=False} respectively, ranked by mean accuracy with a tiebreak on average background rejection. Across the sweep, the gain from enabling GMP is large and consistent (typically $+4$--$5$ points in average background rejection at fixed configuration), while sensitivity to the remaining hyperparameters is small (Section~\ref{sec:ablations}, Appendix~\ref{app:pooling_mode}). The \texttt{patch\_size=10} and \texttt{learned\_pool}/\texttt{mean} configurations used in the main text are within standard error of the best entry in this sweep.

Definitions of each hyperparameter
\begin{itemize}
    \item \textbf{\texttt{use\_pool}} --- whether \texttt{GeometricPooling} ($\eta$-sorted, strided max-pool) is applied between encoder stages; if \texttt{False}, downsampling is replaced by a plain \texttt{Dense} projection.
    \item \textbf{\texttt{ffn\_activation}} --- activation in the position-wise FFN inside each \texttt{PHATBlock}; either \texttt{relu} or \texttt{gelu}.
    \item \textbf{\texttt{patch\_tokenizer\_mode}} --- how a patch $[P, D]$ is collapsed into a single patch token $[D]$:
    \begin{itemize}
        \item \texttt{mean} --- average over the $P$ particles in the patch.
        \item \texttt{max} --- element-wise max over the $P$ particles.
        \item \texttt{flatten\_dense} --- flatten $P \cdot D$ features and pass through a learned \texttt{Dense(D)} layer.
        \item \texttt{learned\_pool} --- softmax-weighted sum, with weights produced per-token by a learned \texttt{Dense(1)}.
    \end{itemize}
    \item \textbf{\texttt{message\_proj}} --- whether the patch-level message is passed through a learned \texttt{Dense(d\_model)} projection before being broadcast back to particle tokens.
    \item \textbf{\texttt{message\_gated}} --- whether each broadcast message is multiplied by a per-token sigmoid gate $\sigma(W x_t)$ (token-conditional gating).
    \item \textbf{\texttt{use\_patch\_messages}} --- whether the block runs patch-to-patch attention (\texttt{PatchAttention} over patch tokens, then broadcast). If \texttt{False}, the block is local-only.
    \item \textbf{\texttt{use\_GMP}} --- whether the \texttt{GeometricMessagePassing} module (depthwise 2D conv on the $\eta$--$\phi$ grid, used as a convolutional position encoder) is applied at the start of each block; corresponds to \texttt{use\_cpe} in the code.
    \item \textbf{\texttt{aggregation}} --- final readout pooling over particle tokens before the classification head; \texttt{mean} or \texttt{max} over the token axis.
    \item \textbf{\texttt{sort\_by}} --- particle ordering used to build patches: \texttt{pt} orders by transverse momentum (descending); \texttt{morton} orders by Morton / Z-order code over the $(\eta, \phi)$ grid so spatially nearby particles fall in the same patch.
\end{itemize}

\begin{table*}[!ht]
\setlength{\tabcolsep}{3pt}
\centering
\scriptsize
\begin{tabular}{lccccccccccccc}
\toprule
\textbf{use\_pool} &
\textbf{\makecell{ffn \\ activation}} &
\textbf{\makecell{patch\_tokenizer \\ mode}} &
\textbf{\makecell{message \\ proj}} &
\textbf{\makecell{message \\ gated}} &
\textbf{\makecell{Test \\ Acc (\%)}} &
\textbf{\makecell{ROC \\ AUC}} &
\textbf{\makecell{Bg Rej \\ W}} &
\textbf{\makecell{Bg Rej \\ Z}} &
\textbf{\makecell{Bg Rej \\ t}} &
\textbf{\makecell{Avg \\ Bg Rej}} &
\textbf{\makecell{\# \\ Params}} &
\textbf{FLOPs} \\
\midrule
False & gelu & mean & True & False & 81.89 $\pm$ 0.03 & 0.962 & 96.3 $\pm$ 1.2 & 92.7 $\pm$ 1.2 & 23.2 $\pm$ 0.2 & 70.7 $\pm$ 0.3 & 6,405 & 968,736 \\
False & gelu & learned\_pool & True & True & 81.88 $\pm$ 0.06 & 0.962 & 96.9 $\pm$ 5.3 & 95.9 $\pm$ 0.9 & 22.9 $\pm$ 0.4 & 71.9 $\pm$ 2.2 & 6,694 & 946,774 \\
True & gelu & mean & True & False & 81.87 $\pm$ 0.03 & 0.962 & 99.9 $\pm$ 1.8 & 93.2 $\pm$ 0.6 & 23.2 $\pm$ 0.1 & 72.1 $\pm$ 0.5 & 6,405 & 968,736 \\
True & gelu & mean & False & False & 81.85 $\pm$ 0.05 & 0.962 & 98.4 $\pm$ 2.4 & 94.8 $\pm$ 3.5 & 22.9 $\pm$ 0.4 & 72.0 $\pm$ 1.2 & 6,133 & 965,568 \\
True & gelu & learned\_pool & False & False & 81.83 $\pm$ 0.02 & 0.962 & 93.3 $\pm$ 3.3 & 94.3 $\pm$ 1.8 & 22.9 $\pm$ 0.0 & 70.2 $\pm$ 1.0 & 6,150 & 973,574 \\
True & relu & mean & True & True & 81.80 $\pm$ 0.08 & 0.962 & 94.4 $\pm$ 2.6 & 97.0 $\pm$ 1.8 & 22.8 $\pm$ 0.4 & 71.4 $\pm$ 1.0 & 6,677 & 900,368 \\
False & relu & mean & True & False & 81.80 $\pm$ 0.12 & 0.962 & 97.9 $\pm$ 0.8 & 92.9 $\pm$ 2.8 & 23.0 $\pm$ 0.7 & 71.3 $\pm$ 0.8 & 6,405 & 930,336 \\
True & gelu & mean & True & True & 81.79 $\pm$ 0.10 & 0.962 & 95.8 $\pm$ 1.6 & 93.4 $\pm$ 2.1 & 22.7 $\pm$ 0.9 & 70.6 $\pm$ 0.7 & 6,677 & 938,768 \\
True & gelu & learned\_pool & True & False & 81.78 $\pm$ 0.08 & 0.962 & 95.4 $\pm$ 3.6 & 98.4 $\pm$ 2.6 & 22.8 $\pm$ 0.5 & 72.2 $\pm$ 0.9 & 6,422 & 976,742 \\
True & gelu & learned\_pool & True & True & 81.78 $\pm$ 0.06 & 0.962 & 95.5 $\pm$ 2.9 & 92.5 $\pm$ 4.8 & 22.6 $\pm$ 0.5 & 70.2 $\pm$ 2.0 & 6,694 & 946,774 \\
\bottomrule
\end{tabular}
\caption{Hyperparameter sweep results: top 10 configurations ranked by mean test accuracy (tiebreak by average background rejection), with use\_patch\_messages = True. All ROC AUC uncertainties are smaller than 0.001, hence not reported here.}
\label{tab:ppt_sweep_top10_no_agg_GMP_auc3_bgrej1}
\end{table*}

\begin{table*}[!ht]
\setlength{\tabcolsep}{3pt}
\centering
\scriptsize
\begin{tabular}{lccccccccccccc}
\toprule
\textbf{use\_pool} &
\textbf{\makecell{ffn \\ activation}} &
\textbf{\makecell{patch\_tokenizer \\ mode}} &
\textbf{\makecell{message \\ proj}} &
\textbf{\makecell{message \\ gated}} &
\textbf{\makecell{Test \\ Acc (\%)}} &
\textbf{\makecell{ROC \\ AUC}} &
\textbf{\makecell{Bg Rej \\ W}} &
\textbf{\makecell{Bg Rej \\ Z}} &
\textbf{\makecell{Bg Rej \\ t}} &
\textbf{\makecell{Avg \\ Bg Rej}} &
\textbf{\makecell{\# \\ Params}} &
\textbf{FLOPs} \\
\midrule
False & gelu & mean & True & False & 81.23 $\pm$ 0.03 & 0.959 & 88.8 $\pm$ 3.4 & 86.5 $\pm$ 1.5 & 19.1 $\pm$ 0.4 & 64.8 $\pm$ 0.7 & 5,061 & 864,484 \\
False & gelu & learned\_pool & True & True & 81.28 $\pm$ 0.03 & 0.960 & 89.4 $\pm$ 1.4 & 87.6 $\pm$ 4.5 & 19.5 $\pm$ 0.3 & 65.5 $\pm$ 1.9 & 5,350 & 842,522 \\
True & gelu & mean & True & False & 81.34 $\pm$ 0.04 & 0.960 & 87.1 $\pm$ 3.4 & 87.2 $\pm$ 2.1 & 19.7 $\pm$ 0.4 & 64.7 $\pm$ 1.6 & 5,061 & 864,484 \\
True & gelu & mean & False & False & 81.31 $\pm$ 0.05 & 0.960 & 87.3 $\pm$ 3.5 & 88.5 $\pm$ 4.4 & 19.4 $\pm$ 0.0 & 65.1 $\pm$ 0.5 & 4,789 & 861,316 \\
True & gelu & learned\_pool & False & False & 81.28 $\pm$ 0.14 & 0.960 & 88.1 $\pm$ 3.6 & 87.3 $\pm$ 4.5 & 19.7 $\pm$ 0.7 & 65.0 $\pm$ 2.7 & 4,806 & 869,322 \\
True & relu & mean & True & True & 81.33 $\pm$ 0.16 & 0.960 & 88.7 $\pm$ 3.3 & 86.8 $\pm$ 1.4 & 19.9 $\pm$ 0.8 & 65.1 $\pm$ 1.5 & 5,333 & 796,116 \\
False & relu & mean & True & False & 81.35 $\pm$ 0.09 & 0.960 & 86.5 $\pm$ 3.4 & 87.9 $\pm$ 0.8 & 19.7 $\pm$ 0.6 & 64.7 $\pm$ 1.5 & 5,061 & 826,084 \\
True & gelu & mean & True & True & 81.39 $\pm$ 0.13 & 0.960 & 91.8 $\pm$ 3.5 & 90.3 $\pm$ 4.3 & 20.0 $\pm$ 0.2 & 67.4 $\pm$ 2.6 & 5,333 & 834,516 \\
True & gelu & learned\_pool & True & False & 81.35 $\pm$ 0.16 & 0.960 & 92.0 $\pm$ 5.2 & 86.5 $\pm$ 5.1 & 19.7 $\pm$ 0.3 & 66.1 $\pm$ 3.2 & 5,078 & 872,490 \\
True & gelu & learned\_pool & True & True & 81.27 $\pm$ 0.02 & 0.960 & 89.5 $\pm$ 2.6 & 87.1 $\pm$ 2.6 & 19.4 $\pm$ 0.4 & 65.3 $\pm$ 0.3 & 5,350 & 842,522 \\
\bottomrule
\end{tabular}
\caption{Top 10 models for the ablation with use\_patch\_messages = True and use\_GMP = False. All ROC AUC uncertainties are smaller than 0.001, hence not reported here.}
\label{tab:ppt_ablation_no_GMP_top10_auc3_bgrej1_noagg}
\end{table*}

\begin{table*}[!ht]
\setlength{\tabcolsep}{6pt}
\centering
\scriptsize
\begin{tabular}{lcccccccccccc}
\toprule
\textbf{use\_pool} &
\textbf{\makecell{aggre- \\ gation}} &
\textbf{\makecell{use \\ GMP}} &
\textbf{\makecell{ffn \\ activation}} &
\textbf{\makecell{Test \\ Acc (\%)}} &
\textbf{\makecell{ROC \\ AUC}} &
\textbf{\makecell{Bg Rej \\ W}} &
\textbf{\makecell{Bg Rej \\ Z}} &
\textbf{\makecell{Bg Rej \\ t}} &
\textbf{\makecell{Avg \\ Bg Rej}} &
\textbf{\makecell{\# \\ Params}} &
\textbf{FLOPs} \\
\midrule
False & mean & True & gelu & 81.76 $\pm$ 0.02 & 0.961 & 92.3 $\pm$ 1.7 & 94.4 $\pm$ 3.5 & 22.2 $\pm$ 0.4 & 69.7 $\pm$ 1.2 & 5,045 & 924,877 \\
True & mean & True & gelu & 81.71 $\pm$ 0.03 & 0.961 & 94.2 $\pm$ 1.5 & 89.3 $\pm$ 0.7 & 22.1 $\pm$ 0.4 & 68.5 $\pm$ 0.7 & 5,045 & 924,877 \\
True & mean & True & relu & 81.73 $\pm$ 0.07 & 0.961 & 94.2 $\pm$ 3.5 & 89.2 $\pm$ 2.7 & 22.3 $\pm$ 0.2 & 68.6 $\pm$ 1.0 & 5,045 & 886,477 \\
False & mean & True & relu & 81.68 $\pm$ 0.04 & 0.961 & 93.4 $\pm$ 3.5 & 89.5 $\pm$ 3.8 & 21.9 $\pm$ 0.4 & 68.2 $\pm$ 1.6 & 5,045 & 886,477 \\
True & max & True & gelu & 81.44 $\pm$ 0.12 & 0.960 & 86.1 $\pm$ 2.4 & 84.5 $\pm$ 2.0 & 21.4 $\pm$ 0.5 & 64.0 $\pm$ 1.6 & 5,045 & 922,477 \\
False & max & True & relu & 81.28 $\pm$ 0.15 & 0.960 & 88.6 $\pm$ 0.9 & 82.6 $\pm$ 4.5 & 20.8 $\pm$ 0.8 & 64.0 $\pm$ 1.9 & 5,045 & 884,077 \\
False & max & True & gelu & 81.45 $\pm$ 0.10 & 0.960 & 90.1 $\pm$ 0.7 & 87.9 $\pm$ 2.5 & 21.0 $\pm$ 0.2 & 66.4 $\pm$ 1.0 & 5,045 & 922,477 \\
True & max & True & relu & 81.35 $\pm$ 0.17 & 0.960 & 86.1 $\pm$ 2.5 & 84.2 $\pm$ 5.6 & 20.9 $\pm$ 0.7 & 63.7 $\pm$ 2.6 & 5,045 & 884,077 \\
True & max & False & gelu & 80.97 $\pm$ 0.04 & 0.958 & 85.2 $\pm$ 2.4 & 81.5 $\pm$ 3.3 & 18.7 $\pm$ 0.3 & 61.8 $\pm$ 1.9 & 3,701 & 818,225 \\
True & max & False & relu & 80.84 $\pm$ 0.10 & 0.958 & 85.1 $\pm$ 4.6 & 80.1 $\pm$ 2.5 & 17.9 $\pm$ 0.3 & 61.1 $\pm$ 0.8 & 3,701 & 779,825 \\
\bottomrule
\end{tabular}
\caption{Top 10 unique configurations from runs with use\_patch\_messages = False. All ROC AUC uncertainties are smaller than 0.001, hence not reported here.}
\label{tab:ppt_unique_configs_no_patch_messages_auc3_bgrej1}
\end{table*}

\begin{table*}[!ht]
\setlength{\tabcolsep}{2pt}
\centering
\scriptsize
\begin{tabular}{lcccccccccccc}
\toprule
\textbf{use\_pool} &
\textbf{\makecell{ffn \\ activation}} &
\textbf{\makecell{patch\_tokenizer \\ mode}} &
\textbf{\makecell{message \\ proj}} &
\textbf{\makecell{message \\ gated}} &
\textbf{\makecell{Test \\ Acc (\%)}} &
\textbf{\makecell{ROC \\ AUC}} &
\textbf{\makecell{Bg Rej \\ W}} &
\textbf{\makecell{Bg Rej \\ Z}} &
\textbf{\makecell{Bg Rej \\ t}} &
\textbf{\makecell{Avg \\ Bg Rej}} &
\textbf{\makecell{\# \\ Params}} &
\textbf{FLOPs} \\
\midrule
False & gelu & mean         & True  & False & 81.81 $\pm$ 0.09 & 0.962 & 96.3 $\pm$ 3.0 & 95.4 $\pm$ 2.1 & 23.4 $\pm$ 0.9 & 71.7 $\pm$ 1.5 & 6,405 & 968,736 \\
False & gelu & learned\_pool & True  & True  & 81.83 $\pm$ 0.09 & 0.962 & 95.6 $\pm$ 4.2 & 96.9 $\pm$ 2.2 & 22.9 $\pm$ 0.3 & 71.8 $\pm$ 2.2 & 6,694 & 946,774 \\
True  & gelu & mean         & True  & False & 81.84 $\pm$ 0.10 & 0.962 & 93.9 $\pm$ 4.7 & 95.4 $\pm$ 1.3 & 22.7 $\pm$ 0.7 & 70.7 $\pm$ 1.8 & 6,405 & 968,736 \\
True  & gelu & mean         & False & False & 81.73 $\pm$ 0.06 & 0.962 & 95.0 $\pm$ 3.1 & 93.3 $\pm$ 3.5 & 22.8 $\pm$ 0.4 & 70.4 $\pm$ 0.6 & 6,133 & 965,568 \\
True  & gelu & learned\_pool & False & False & 81.88 $\pm$ 0.05 & 0.962 & 94.6 $\pm$ 1.6 & 95.9 $\pm$ 3.1 & 23.4 $\pm$ 0.5 & 71.3 $\pm$ 1.0 & 6,150 & 973,574 \\
True  & relu & mean         & True  & True  & 81.79 $\pm$ 0.07 & 0.962 & 93.7 $\pm$ 3.7 & 95.4 $\pm$ 3.0 & 22.9 $\pm$ 0.1 & 70.7 $\pm$ 1.0 & 6,677 & 900,368 \\
False & relu & mean         & True  & False & 81.85 $\pm$ 0.14 & 0.962 & 94.8 $\pm$ 5.9 & 94.0 $\pm$ 4.0 & 22.9 $\pm$ 0.5 & 70.6 $\pm$ 3.3 & 6,405 & 930,336 \\
True  & gelu & mean         & True  & True  & 81.77 $\pm$ 0.03 & 0.962 & 93.7 $\pm$ 1.6 & 92.5 $\pm$ 1.0 & 22.6 $\pm$ 0.4 & 69.6 $\pm$ 0.8 & 6,677 & 938,768 \\
True  & gelu & learned\_pool & True  & False & 81.80 $\pm$ 0.09 & 0.962 & 99.6 $\pm$ 3.6 & 94.3 $\pm$ 1.9 & 22.4 $\pm$ 0.0 & 72.1 $\pm$ 1.5 & 6,422 & 976,742 \\
True  & gelu & learned\_pool & True  & True  & 81.78 $\pm$ 0.04 & 0.962 & 94.3 $\pm$ 1.2 & 93.3 $\pm$ 0.7 & 22.8 $\pm$ 0.6 & 70.1 $\pm$ 0.4 & 6,694 & 946,774 \\
\bottomrule
\end{tabular}
\caption{Sorting ablation results (top 10) with \texttt{sort\_by = pt} and \texttt{use\_patch\_messages = True}. All ROC AUC uncertainties are smaller than 0.001, hence not reported here.}
\label{tab:sorting_ablation_pt_auc3_bgrej1_noagg}
\end{table*}

\begin{table*}[!ht]
\setlength{\tabcolsep}{2pt}
\centering
\scriptsize
\begin{tabular}{lcccccccccccc}
\toprule
\textbf{use\_pool} &
\textbf{\makecell{ffn \\ activation}} &
\textbf{\makecell{patch\_tokenizer \\ mode}} &
\textbf{\makecell{message \\ proj}} &
\textbf{\makecell{message \\ gated}} &
\textbf{\makecell{Test \\ Acc (\%)}} &
\textbf{\makecell{ROC \\ AUC}} &
\textbf{\makecell{Bg Rej \\ W}} &
\textbf{\makecell{Bg Rej \\ Z}} &
\textbf{\makecell{Bg Rej \\ t}} &
\textbf{\makecell{Avg \\ Bg Rej}} &
\textbf{\makecell{\# \\ Params}} &
\textbf{FLOPs} \\
\midrule
False & gelu & mean         & True  & False & 81.84 $\pm$ 0.08 & 0.962 & 96.4 $\pm$ 1.9 & 97.4 $\pm$ 3.2 & 23.0 $\pm$ 0.3 & 72.3 $\pm$ 1.4 & 6,405 & 968,736 \\
False & gelu & learned\_pool & True  & True  & 81.80 $\pm$ 0.03 & 0.962 & 95.4 $\pm$ 2.0 & 91.8 $\pm$ 1.4 & 23.0 $\pm$ 0.4 & 70.1 $\pm$ 0.8 & 6,694 & 946,774 \\
True  & gelu & mean         & True  & False & 81.82 $\pm$ 0.09 & 0.962 & 96.9 $\pm$ 0.6 & 91.2 $\pm$ 1.1 & 23.0 $\pm$ 0.5 & 70.4 $\pm$ 0.4 & 6,405 & 968,736 \\
True  & gelu & mean         & False & False & 81.82 $\pm$ 0.05 & 0.962 & 95.7 $\pm$ 6.0 & 94.3 $\pm$ 2.4 & 23.3 $\pm$ 0.3 & 71.1 $\pm$ 2.1 & 6,133 & 965,568 \\
True  & gelu & learned\_pool & False & False & 81.76 $\pm$ 0.04 & 0.962 & 95.3 $\pm$ 0.9 & 93.8 $\pm$ 2.2 & 22.9 $\pm$ 0.3 & 70.7 $\pm$ 0.9 & 6,150 & 973,574 \\
True  & relu & mean         & True  & True  & 81.78 $\pm$ 0.05 & 0.962 & 94.8 $\pm$ 3.0 & 93.5 $\pm$ 2.0 & 22.8 $\pm$ 0.4 & 70.4 $\pm$ 0.5 & 6,677 & 900,368 \\
False & relu & mean         & True  & False & 81.76 $\pm$ 0.04 & 0.962 & 93.9 $\pm$ 2.4 & 92.5 $\pm$ 3.3 & 22.9 $\pm$ 0.5 & 69.7 $\pm$ 1.8 & 6,405 & 930,336 \\
True  & gelu & mean         & True  & True  & 81.73 $\pm$ 0.05 & 0.962 & 98.5 $\pm$ 1.5 & 92.9 $\pm$ 3.8 & 22.7 $\pm$ 0.2 & 71.3 $\pm$ 1.4 & 6,677 & 938,768 \\
True  & gelu & learned\_pool & True  & False & 81.85 $\pm$ 0.07 & 0.962 & 96.3 $\pm$ 2.9 & 94.0 $\pm$ 2.0 & 23.3 $\pm$ 0.5 & 71.2 $\pm$ 1.2 & 6,422 & 976,742 \\
True  & gelu & learned\_pool & True  & True  & 81.80 $\pm$ 0.14 & 0.962 & 94.7 $\pm$ 5.3 & 95.8 $\pm$ 2.0 & 23.2 $\pm$ 0.5 & 71.2 $\pm$ 2.3 & 6,694 & 946,774 \\
\bottomrule
\end{tabular}
\caption{Top 10 models for the ablation with \texttt{use\_patch\_messages = True} and \texttt{use\_GMP = False}. All ROC AUC uncertainties are smaller than 0.001 and are not reported here.}
\label{tab:ppt_ablation_no_GMP_top10_auc3_bgrej1_noagg}
\end{table*}

\newpage
\section{Heuristic Coverage Model for Patch Attention}
\label{app:coverage-model}
This appendix provides a simple probabilistic model for how GMP and hierarchical patch-token
attention increase the chance that a relevant particle-particle interaction is represented. The model is
intended as a coverage heuristic rather than a formal expressivity guarantee.

Consider a relevant particle pair $(i,j)$ in a jet with $N$ particles and patch size $P$. Under an
exchangeable patch-assignment model, the probability that $i$ and $j$ appear in the same local patch is
\begin{equation}
q=\frac{P-1}{N-1}.
\end{equation}
Thus patch attention alone exposes the pair to exact local attention with probability
\begin{equation}
p_{\mathrm{patch}}=q.
\end{equation}

GMP increases this probability by spreading each particle's information over an effective local support.
Let $s$ denote the effective number of particle tokens that carry information about a given particle
after GMP. Under an independence approximation, a pair $(i,j)$ has approximately $s^2$ possible
token-level bridge opportunities. The probability that at least one such bridge is exposed within a
local patch is therefore
\begin{equation}
p_{\mathrm{GMP}}
\approx
1-(1-q)^{s^2}.
\end{equation}

The global patch-token stage provides an additional recovery path for interactions missed locally.
Let $\rho$ denote the effective probability that a particle's information is preserved usefully in its
patch token. Since patch-token attention is global over the $M=\lceil N/P\rceil$ patch summaries, a
missed pair can still be recovered if useful information about both particles survives patch pooling.
This gives
\begin{equation}
p_{\mathrm{GMP+global}}
\approx
p_{\mathrm{GMP}}+(1-p_{\mathrm{GMP}})\rho^2
=
1-(1-q)^{s^2}(1-\rho^2).
\end{equation}

For the HLS4ML setting used in our main experiments, $N=150$ and $P=10$, so
\begin{equation}
q=\frac{9}{149}\approx 0.060.
\end{equation}
Using $s=4$ as a conservative same-cell GMP support estimate at grid spacing $\delta=0.2$ gives
\begin{equation}
p_{\mathrm{GMP}}
\approx
1-(1-0.060)^{16}
\approx
0.63.
\end{equation}
The total coverage estimate is therefore
\begin{equation}
p_{\mathrm{GMP+global}}
\approx
0.63+0.37\rho^2.
\end{equation}

Under this model, the effective coverage of a relevant particle-particle interaction changes as
\begin{equation}
6\%
\quad\longrightarrow\quad
63\%
\quad\longrightarrow\quad
63\%+37\%\rho^2.
\end{equation}
This explains the complementary roles of the three components: patch attention provides exact
local interactions, GMP creates multiple local opportunities for the same information to be exposed,
and global patch-token attention provides a recovery path for information separated across patches.
It also gives an intuition for the observed robustness to consistent particle orderings since each
particle is represented through multiple effective token-level connections, the model depends less on
any single connection. 

\section{Training and Testing Setup}
\label{app:training}
All models in the main results are trained with categorical cross-entropy loss using the Adam optimizer with default parameters ($\beta_1{=}0.9$, $\beta_2{=}0.999$). We use a batch-size scheduler that increases the batch size during training to stabilize the loss landscape near convergence. Models are trained for up to 200 epochs with early stopping on validation loss (patience 20). We use the official training/validation/test splits provided with each dataset (\textsc{hls4ml}, JetClass, Top Tagging) and a 20\% validation split sampled from the training set for Quark--Gluon. For all transformer-based models we use four attention heads, hidden dimension 16 (for \textsc{hls4ml}, Top Tagging, Quark--Gluon) or 32 (for JetClass), and a single attention layer to match the trigger-resource regime. PHAT-JeT uses a patch size of $P=10$ and a GMP grid spacing of $\delta=0.2$ throughout, unless otherwise noted. All input features are normalized as described in Appendix~\ref{app:dataset}. We report mean $\pm$ standard deviation across $n{=}3$ random seeds.

\section{Baseline Models}
\label{app:baselines}
Table~\ref{tab:sorting_non_equivariant_baselines_grouped} shows that non-equivariant baseline models see a performance improvement upon switching from the default sorting scheme to $\kT$ sorting.

\begin{table}[h!]
\centering
\small
\setlength{\tabcolsep}{4pt}
\begin{tabular}{lccc}
\toprule
\textbf{Model (Sort)} &
\textbf{\makecell{Acc \\ (\%)}} &
\textbf{\makecell{ROC \\ AUC}} &
\textbf{\makecell{Bkg Rej}} \\
\toprule
\textbf{Linformer ($\kT$ sorted)} &
\textbf{81.00 $\pm$ 0.08} & \textbf{0.959} & \textbf{38.4 $\pm$ 0.5} \\
Linformer ($\pT$ sorted) &
79.90 $\pm$ 0.00 & 0.955 & 28.1 $\pm$ 0.6 \\
\midrule
\textbf{SAL-T ($\kT$ sorted)} &
\textbf{81.18 $\pm$ 0.03} & \textbf{0.959} & \textbf{40.8 $\pm$ 0.6} \\
SAL-T ($\pT$ sorted) &
78.82 $\pm$ 0.01 & 0.950 & 23.5 $\pm$ 2.1 \\
\midrule
\textbf{Point Transformer v3 ($\kT$ sorted)} &
\textbf{81.37 $\pm$ 0.06} & \textbf{0.960} & \textbf{66.8 $\pm$ 0.2} \\
Point Transformer v3 ($\pT$ sorted) &
81.32 $\pm$ 0.00 & 0.960 & 65.8 $\pm$ 1.4 \\
Point Transformer v3 (Morton ordering) &
80.99 $\pm$ 0.15 & 0.958 & 60.2 $\pm$ 1.4 \\
\bottomrule
\end{tabular}
\caption{Performance of non-permutation-equivariant baseline models on the \textsc{hls4ml} dataset. These models benefit from $\kT$ sorting, which is the default sorting scheme we use in the reported benchmarks in this manuscript. \textbf{Bkg Rej} is $1/\mathrm{FPR}@0.8\,\mathrm{TPR}$ (higher is better).}
\label{tab:sorting_non_equivariant_baselines_grouped}
\end{table}

\section{GMP Grid Spacing Sensitivity}
\label{app:grid_spacing}
The grid spacing $\delta$ controls the angular resolution of the GMP module. Table~\ref{tab:grid_spacing} sweeps $\delta$ for the full PHAT-JeT model on \textsc{hls4ml}. Performance is stable across $\delta=0.1$--$0.2$ and degrades modestly at $\delta=0.3$ where distinct angular substructures merge into the same cell. Very fine grids ($\delta=0.05$) do not help because cells become sparse and the depthwise convolution has little to mix. The stable range corresponds to roughly 8--16 cells across a single jet cone, which is consistent with the angular scale of multi-prong heavy-particle decays.

\begin{table}[h!]
\centering
\small
\setlength{\tabcolsep}{6pt}
\begin{tabular}{lcc}
\toprule
$\delta$ & \textbf{Acc (\%)} & \textbf{Avg Bg Rej} \\
\midrule
0.1 & 81.83 $\pm$ 0.03 & 70.0 $\pm$ 2.6 \\
0.2 (default) & 81.78 $\pm$ 0.01 & 70.3 $\pm$ 0.7 \\
0.3 & 81.68 $\pm$ 0.03 & 68.2 $\pm$ 0.7 \\
\bottomrule
\end{tabular}
\caption{GMP grid spacing $\delta$ sweep for PHAT-JeT on \textsc{hls4ml}. ROC AUC differences are below 0.001.}
\label{tab:grid_spacing}
\end{table}

We additionally tested scaling the GMP grid coordinates by particle $\pT$ (so that high-$\pT$ particles influence a wider angular footprint), which consistently underperformed raw $(\eta,\phi)$ coordinates. We attribute this to angular position being the more informative geometric axis for capturing multi-prong jet substructure.

\paragraph{Sum-pooling and multiplicity sensitivity.} A natural concern is that the per-cell sum pooling in Eq.~(\ref{eq:gmp_sum}) (where multiple particles in the same grid cell are summed) could introduce a systematic bias for jets with very different multiplicities. In the \textsc{hls4ml} dataset all jets are zero-padded to a fixed multiplicity of $N=150$, so the upper bound on the per-cell sum is the same across jets, but the \emph{effective} multiplicity (number of non-zero particles) does vary by class. Empirically the per-class effective multiplicities cluster around: $q$ (light quark) $\sim$23, $g$ (gluon) $\sim$30, $W$ $\sim$45, $Z$ $\sim$48, top $\sim$70 -- a factor of $\sim$3 spread between the lowest and highest classes. In principle, this could shift the magnitude of $G_{u,v}$ for top jets relative to quark jets and bias the convolution output. Two design choices mitigate this. First, the GMP output $Y$ in Eq.~(\ref{eq:gmp_residual}) is added to a layer-normalized residual stream rather than used directly downstream, so the network can absorb a class-dependent magnitude shift through its standard normalization machinery. Second, the depthwise convolution operating on $G$ has a small kernel (3$\times$3) and the residual blends with per-particle features that already carry $\pT$-dependent information, so it does not need to recover absolute multiplicity from the sum alone. Empirically, the GMP gain is approximately uniform across jet classes: per-class background rejection improvements are within 1.5 standard deviations for $q$, $g$, $W$, $Z$, and $t$ (computed from Tables~\ref{tab:ppt_sweep_top10_no_agg_GMP_auc3_bgrej1} and~\ref{tab:ppt_ablation_no_GMP_top10_auc3_bgrej1_noagg}), which is what we would expect if multiplicity-induced bias were not a dominant effect. We note this as a possible concern under a different padding scheme or for datasets where the multiplicity distribution is much wider than \textsc{hls4ml}.

\section{FLOPs-Matched Baselines}
\label{app:scaled_baselines}
Table~\ref{tab:scaled_baselines} reports performance when every baseline is scaled (in hidden dimension) so its FLOP count matches PHAT-JeT's, isolating the contribution of the architecture from the total compute budget. Linformer and SAL-T are additionally augmented with the GMP module, since GMP is shown to transfer across architectures (Table~\ref{tab:ablations_big_blocks_desc}). PHAT-JeT remains the strongest model under this strict matched-compute comparison.

\begin{table}[h!]
\centering
\small
\setlength{\tabcolsep}{4pt}
\begin{tabular}{lcccc}
\toprule
\textbf{Model} & \textbf{Acc (\%)} & \textbf{Avg Bg Rej} & \textbf{\# Params (K)} & \textbf{FLOPs (M)} \\
\midrule
\textbf{PHAT-JeT}        & $\mathbf{81.80 \pm 0.02}$ & $\mathbf{71.6 \pm 0.6}$ & 6.7  & 1.31 \\
Linformer + GMP          & $81.55 \pm 0.03$          & $67.0 \pm 0.7$          & 10.1 & 1.36 \\
SAL-T + GMP              & $81.52 \pm 0.06$          & $65.7 \pm 1.4$          & 5.4  & 1.33 \\
PointTransformer V3      & $80.93 \pm 0.10$          & $61.6 \pm 0.6$          & 22.4 & 1.24 \\
PointNet                 & $75.04 \pm 0.06$          & $17.0 \pm 0.1$          & 9.8  & 1.35 \\
\bottomrule
\end{tabular}
\caption{FLOPs-matched comparison on \textsc{hls4ml}. Each baseline is scaled in hidden dimension to bring its FLOPs into PHAT-JeT's range, and Linformer/SAL-T additionally receive the GMP module, isolating the architectural contribution from the compute budget.}
\label{tab:scaled_baselines}
\end{table}

\section{Patch-Token Pooling Mode}
\label{app:pooling_mode}
The patch-level attention stage requires aggregating each patch into a single token before applying the global multi-head self-attention. Table~\ref{tab:pooling_mode} compares four pooling strategies, all evaluated with GMP enabled. Mean and learned-pool attain comparable accuracy ($81.89\%$ vs.\ $81.88\%$), with mean pooling giving tighter background-rejection variance. Max pooling and Flatten+Dense both trail by approximately 0.4\% accuracy and 4 points in average background rejection. The fact that strictly more expressive pooling strategies do not outperform the simple mean suggests that the mean-pooled patch summary already captures the information relevant for inter-patch communication, which supports the use of plain mean pooling in the final model.

\begin{table}[h!]
\centering
\small
\setlength{\tabcolsep}{6pt}
\begin{tabular}{lcc}
\toprule
\textbf{Pooling Mode} & \textbf{Acc (\%)} & \textbf{Avg Bg Rej} \\
\midrule
Mean (default)   & 81.89 $\pm$ 0.03 & 70.7 $\pm$ 0.3 \\
Learned pool     & 81.88 $\pm$ 0.06 & 71.9 $\pm$ 2.2 \\
Flatten + dense  & 81.54 $\pm$ 0.03 & 66.6 $\pm$ 0.9 \\
Max              & 81.45 $\pm$ 0.10 & 66.4 $\pm$ 1.0 \\
\bottomrule
\end{tabular}
\caption{Patch-token pooling mode comparison on \textsc{hls4ml}, all rows use GMP. Numbers are extracted from the hyperparameter sweep in Appendix~\ref{app:ablations} (Tables~\ref{tab:ppt_sweep_top10_no_agg_GMP_auc3_bgrej1}--\ref{tab:ppt_ablation_no_GMP_top10_auc3_bgrej1_noagg}); the column \texttt{patch\_tokenizer\_mode} in those tables corresponds to ``Pooling Mode'' here, while \texttt{message\_proj} and \texttt{message\_gated} indicate whether the broadcast back to particles passes through a learned linear projection $W_m$ and a learned gating scalar, respectively.}
\label{tab:pooling_mode}
\end{table}

\section{Top and QG Tagging Datasets: Setup}
\label{app:top_qg}
This appendix provides the dataset and preprocessing details accompanying the main-text Table~\ref{tab:performance_top_qg}. The Top Tagging dataset~\cite{Kasieczka2019TopQuark} contains 1.2 million jets for training, 400{,}000 for validation, and 400{,}000 for testing. Signal jets originate from the decay of top quarks, while background jets come from light quarks and gluons in dijet events. The Quark--Gluon dataset~\cite{komiske_2019_3164691} contains 1.8 million jets for training and 200{,}000 for testing. We randomly sample 20\% of the training set for validation. This dataset consists of two classes corresponding to quark-initiated and gluon-initiated jets. We use the same training pipeline as for \textsc{hls4ml} (categorical cross-entropy loss, Adam optimizer, batch-size scheduler), with all baselines constrained to a single attention (or equivalent) layer for resource-matched comparison.

\section{Attention Interpretability}
\label{app:attention}

This appendix gives a qualitative view of what PHAT-JeT attends to, with emphasis on the comparison between $k_T$ sorting and fixed random sorting. 
Our main goal is to show that the learned attention patterns are broadly similar under both orderings, which supports the result in Section~\ref{sec:ablations} that PHAT-JeT performs comparably as long as the train and test ordering are matched. 
This suggests that the patch structure is primarily a computational abstraction, rather than a mechanism that depends on a special physics-motivated ordering.

We compare four attention views throughout:
(i) a standard full-attention Transformer,
(ii) PHAT-JeT local-only attention,
(iii) PHAT-JeT global-only patch-token attention, and
(iv) the combined PHAT-JeT attention.
Unless otherwise stated, the figures show a representative top jet with 69 particles and patch size $P=10$.

\subsection{Per-head attention under $k_T$ and random sorting}

\begin{figure}[!ht]
    \centering
    \includegraphics[width=\linewidth]{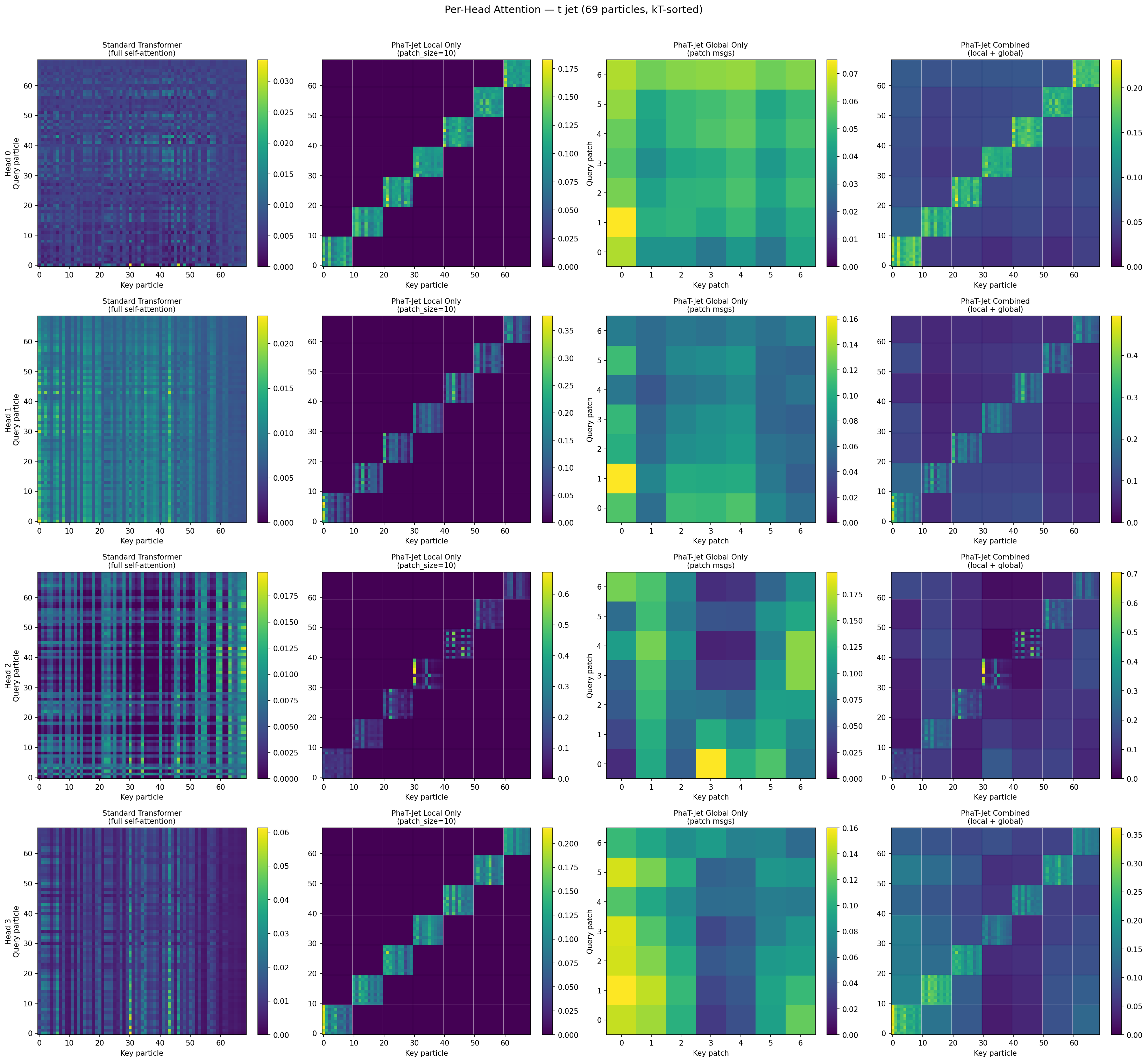}
    \caption{
    Per-head attention on a representative top jet under $k_T$ sorting.
    Rows correspond to attention heads.
    Columns show the standard Transformer, PHAT-JeT local-only attention, PHAT-JeT global-only attention, and combined PHAT-JeT attention.
    }
    \label{fig:attn_per_head_kt}
\end{figure}

Figure~\ref{fig:attn_per_head_kt} shows per-head attention for a representative top jet under $k_T$ sorting. 
As expected, the standard Transformer produces dense full-sequence attention, while PHAT-JeT decomposes attention into a structured local component and a lightweight global component. 
Different heads learn different patterns, but the combined PHAT-JeT attention remains sparse and organized.

\begin{figure}[!ht]
    \centering
    \includegraphics[width=\linewidth]{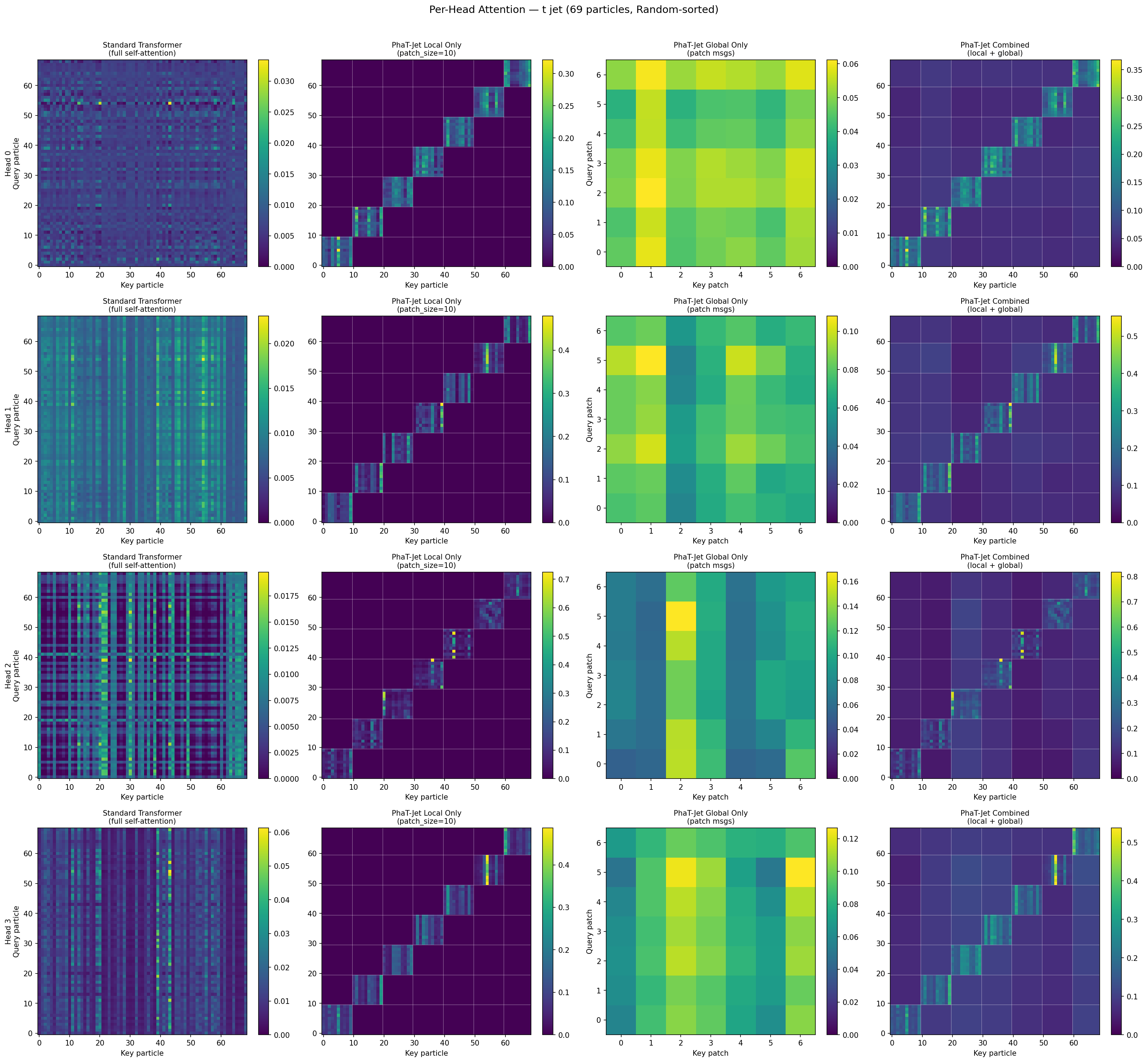}
    \caption{
    Per-head attention on a representative top jet under fixed random sorting.
    The qualitative structure is similar to the $k_T$-sorted case, supporting the observation that PHAT-JeT performs similarly across matched orderings.
    }
    \label{fig:attn_per_head_random}
\end{figure}

Figure~\ref{fig:attn_per_head_random} shows the same visualization for a model trained and evaluated with fixed random ordering. 
The main qualitative picture is unchanged: PHAT-JeT still learns clear local patterns, selective global routing, and a structured combined attention map. 
The similarity between Figures~\ref{fig:attn_per_head_kt} and~\ref{fig:attn_per_head_random} helps explain why matched random ordering performs nearly as well as matched $k_T$ ordering. 
In other words, PHAT-JeT does not rely on $k_T$ sorting to form useful attention patterns; it only requires a consistent ordering so that the learned patch decomposition remains stable.

\subsection{Attention in the $(\eta,\phi)$ plane}

We next project the attention patterns onto the detector plane. 
Particles are plotted in $(\eta,\phi)$ and colored by subjet assignment from declustering into three subjets. 
Solid colored edges denote attention within a subjet, while dashed gray edges denote attention across subjets.

\begin{figure}[!ht]
    \centering
    \includegraphics[width=\linewidth]{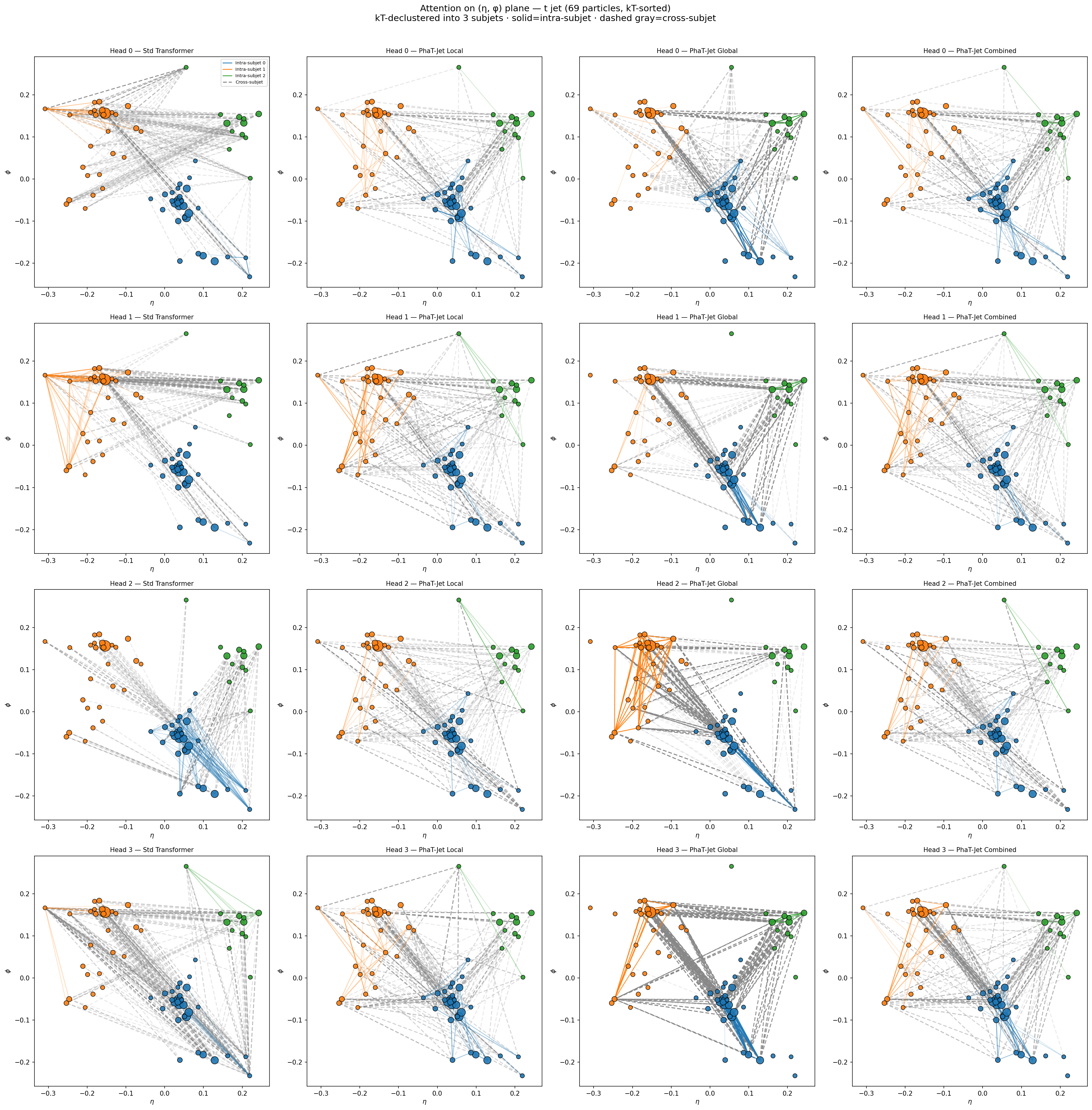}
    \caption{
    Attention projected onto the $(\eta,\phi)$ plane for a representative top jet under $k_T$ sorting.
    Particles are colored by subjet assignment.
    Solid colored edges denote intra-subjet attention, while dashed gray edges denote cross-subjet attention.
    }
    \label{fig:attn_eta_phi_kt}
\end{figure}

Figure~\ref{fig:attn_eta_phi_kt} shows the $k_T$-sorted case. 
The standard Transformer produces many diffuse cross-particle connections, whereas PHAT-JeT yields a more selective pattern. 
The local, global, and combined stages play distinct roles, but the main point is that the model captures structured relations between particles rather than spreading attention uniformly.

\begin{figure}[!ht]
    \centering
    \includegraphics[width=\linewidth]{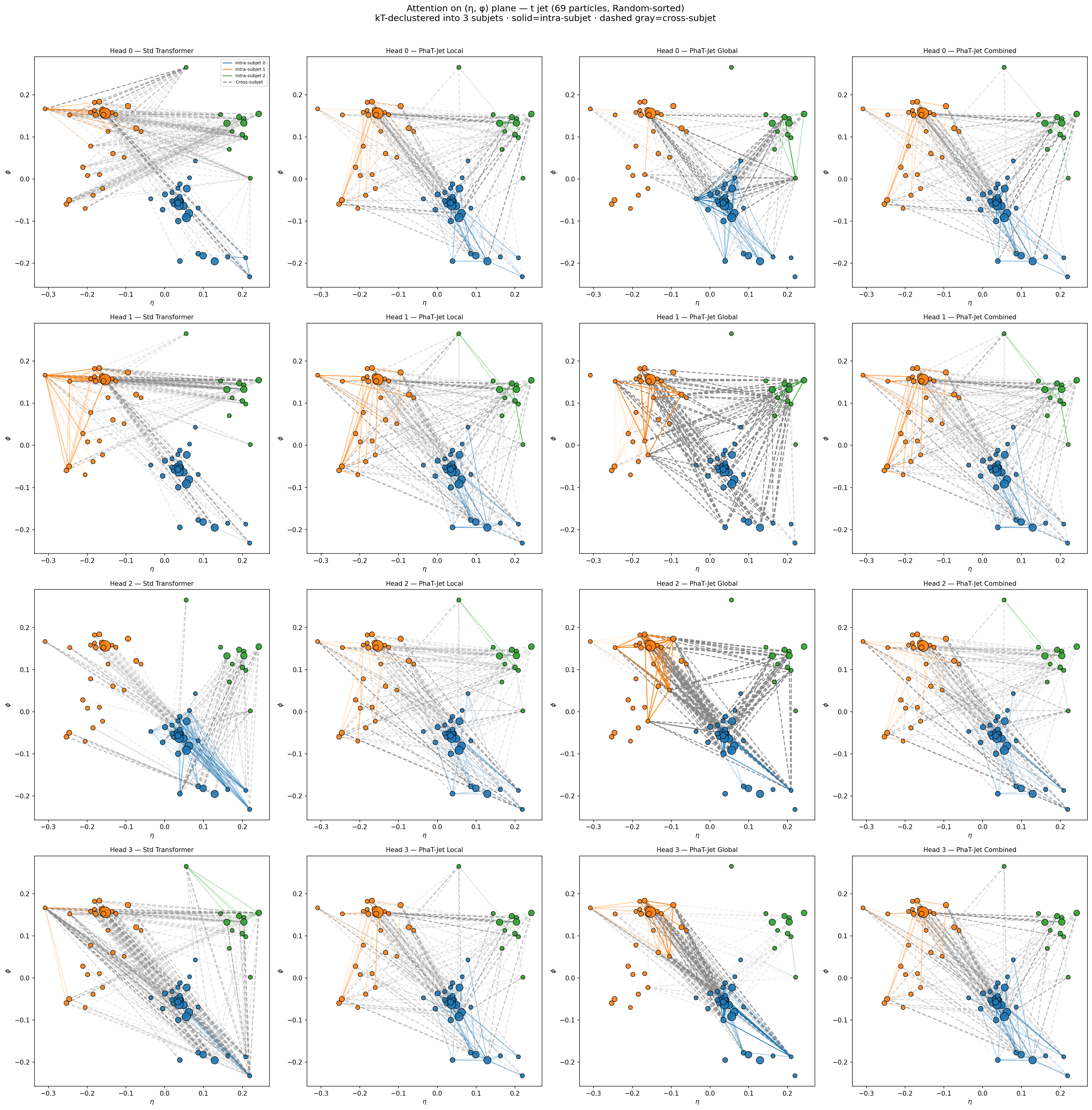}
    \caption{
    Attention projected onto the $(\eta,\phi)$ plane for the same type of jet under fixed random sorting.
    The qualitative structure remains similar to the $k_T$-sorted case, consistent with the comparable performance of matched random and matched $k_T$ orderings.
    }
    \label{fig:attn_eta_phi_random}
\end{figure}

Figure~\ref{fig:attn_eta_phi_random} shows the same detector-plane visualization for fixed random sorting. 
Although the particle sequence is no longer ordered by a physics-motivated rule, the qualitative structure remains similar: PHAT-JeT still learns selective and nontrivial attention patterns in the physical jet plane. 
This again supports the conclusion that the model does not depend strongly on $k_T$ ordering itself. 
Instead, it adapts to whatever ordering it is trained on, provided that the same ordering is used at test time.

\subsection{Head-averaged attention across classes}

\begin{figure}[!ht]
    \centering
    \includegraphics[width=\linewidth]{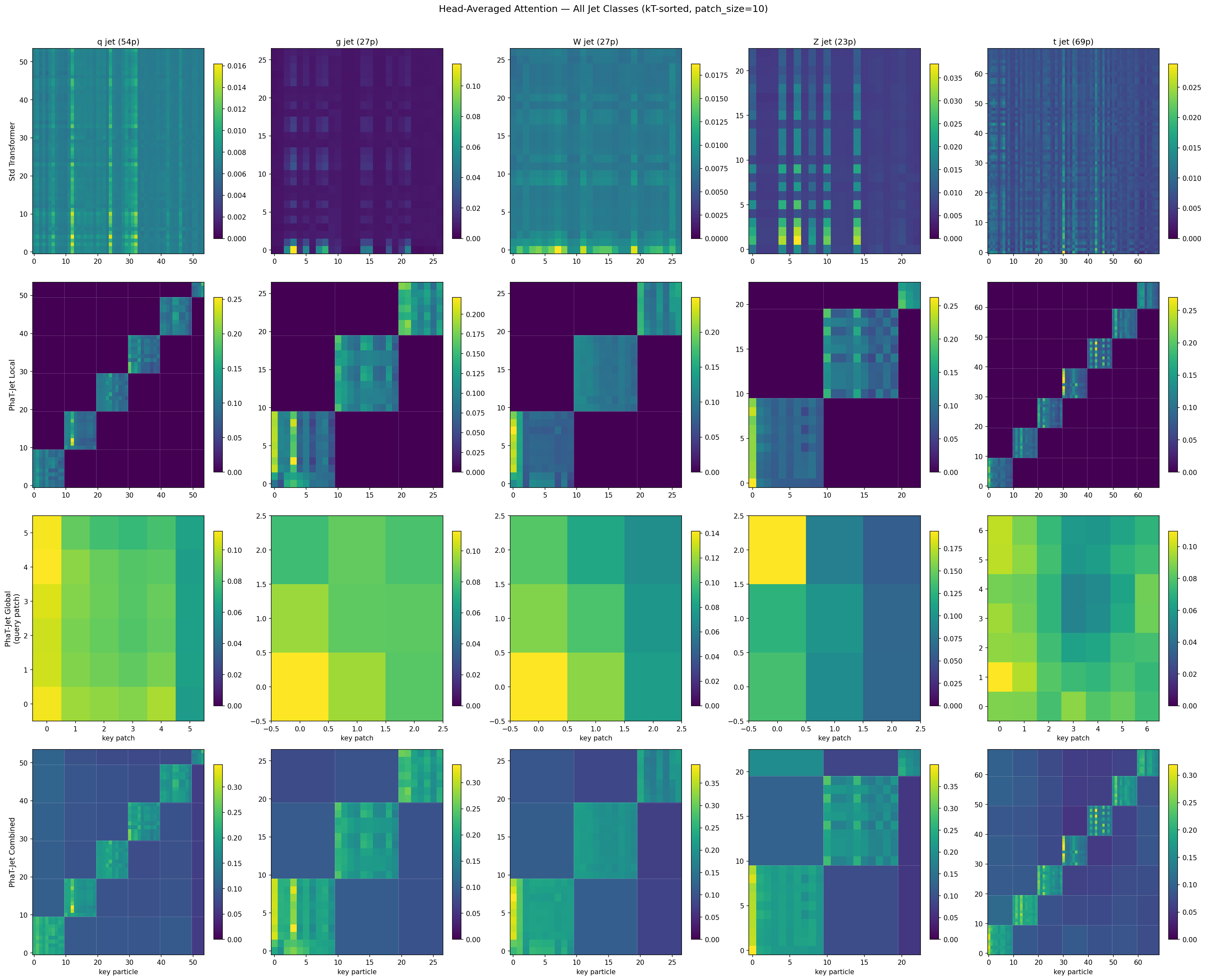}
    \caption{
    Head-averaged attention for representative jets from all five HLS4ML classes.
    Rows correspond to the standard Transformer, PHAT-JeT local-only attention, PHAT-JeT global-only attention, and combined PHAT-JeT attention.
    }
    \label{fig:attn_head_avg}
\end{figure}

To show that this is not specific to a single top jet, Figure~\ref{fig:attn_head_avg} shows head-averaged attention for representative jets from all five HLS4ML classes. 
Across classes, PHAT-JeT consistently displays structured local attention together with non-uniform global patch communication. 
The global stage is clearly not acting as a trivial average: different classes produce different patch-level patterns, indicating that the patch-token mechanism learns class-dependent structure.

\subsection{Summary}

The main conclusion of these visualizations is that PHAT-JeT learns similar attention structure under both $k_T$ sorting and fixed random sorting. 
This provides qualitative support for the quantitative ordering study: PHAT-JeT is robust to the choice of ordering when training and evaluation use the same rule. 
The results therefore justify the interpretation of patching as an efficient attention factorization, rather than as a hard-coded dependence on a particular physics ordering. 
At the same time, the learned local and global attention patterns remain structured and class-dependent, showing that the model is learning nontrivial particle relations rather than relying on a diffuse dense-attention map.

\subsection{Limitations}

This analysis is qualitative and based on representative examples. 
Attention maps should therefore be interpreted as diagnostic visualizations, not as complete causal explanations of the classifier. 
A more complete analysis could quantify similarities between $k_T$-sorted and randomly sorted models over the full test set, for example by comparing attention sparsity, intra- versus inter-subjet fractions, or patch-to-patch attention statistics.
%%%%%%%%%%%%%%%%%%%%%%%%%%%%%%%%%%%%%%%%%%%%%%%%%%%%%%%%%%%%
% NeurIPS Paper Checklist (mandatory; does not count toward page limit)
\newpage
\newpage

\section*{NeurIPS Paper Checklist}

\begin{enumerate}

\item {\bf Claims}
    \item[] Question: Do the main claims made in the abstract and introduction accurately reflect the paper's contributions and scope?
    \item[] Answer: \answerYes{}
    \item[] Justification: The abstract and introduction state the four contributions of the paper (PHAT-JeT architecture, the GMP module, hierarchical patch attention, and the empirical validation), and qualify SOTA claims as ``among resource-constrained models'' to match the experimental scope. Section~\ref{sec:results} reports the corresponding results on the \textsc{hls4ml}, JetClass, Top Tagging, and Quark--Gluon benchmarks, and the limitations of these claims (e.g., FPGA-compatible rather than FPGA-demonstrated) are spelled out in the conclusion and the impact statement.

\item {\bf Limitations}
    \item[] Question: Does the paper discuss the limitations of the work performed by the authors?
    \item[] Answer: \answerYes{}
    \item[] Justification: The conclusion explicitly frames PHAT-JeT as FPGA-compatible rather than FPGA-demonstrated, and notes that end-to-end firmware synthesis is left to future work. Section~\ref{sec:ablations} restricts the ordering claim to robustness under a consistent training ordering rather than full permutation invariance, and table~\ref{tab:order_mismatch} provides the supporting train/test mismatch table. Appendix~\ref{app:grid_spacing} discusses sensitivity to the GMP grid spacing $\delta$ and to potential multiplicity-induced bias in sum pooling. The impact statement additionally notes the simulation-based nature of the training data and the resulting domain-shift considerations.

\item {\bf Theory assumptions and proofs}
    \item[] Question: For each theoretical result, does the paper provide the full set of assumptions and a complete (and correct) proof?
    \item[] Answer: \answerNA{}
    \item[] Justification: The paper does not present formal theorems or proofs; the contribution is an architecture and an empirical study. The complexity analysis ($\mathcal{O}(N\cdot P)$ for local attention, $\mathcal{O}((N/P)^2)$ for the patch-token stage) is informal counting and is stated alongside the equations defining each stage in Section~\ref{sec:method}.

    \item {\bf Experimental result reproducibility}
    \item[] Question: Does the paper fully disclose all the information needed to reproduce the main experimental results of the paper to the extent that it affects the main claims and/or conclusions of the paper (regardless of whether the code and data are provided or not)?
    \item[] Answer: \answerYes{}
    \item[] Justification: Section~\ref{sec:method} fully specifies the architecture (GMP grid construction, patch partitioning, hierarchical attention, and the equations defining each stage). Appendix~\ref{app:training} reports the optimizer (Adam, default $\beta$), batch-size scheduler, early stopping, hidden dimensions per dataset, number of attention heads, patch size $P=10$, and grid spacing $\delta=0.2$. Appendix~\ref{app:dataset} documents the datasets and preprocessing. All experiments use publicly available benchmarks (\textsc{hls4ml}, JetClass, Top Tagging, Quark--Gluon) under their standard splits, and an anonymous code release accompanies the submission (link in the abstract).

\item {\bf Open access to data and code}
    \item[] Question: Does the paper provide open access to the data and code, with sufficient instructions to faithfully reproduce the main experimental results, as described in supplemental material?
    \item[] Answer: \answerYes{}
    \item[] Justification: An anonymized code release is provided at the URL listed in the abstract, including training scripts, model definitions, and configuration files for all experiments. All four datasets used (\textsc{hls4ml}, JetClass, Top Tagging, Quark--Gluon) are publicly available from their original sources, and the README in the code release documents the download and preprocessing pipeline.

\item {\bf Experimental setting/details}
    \item[] Question: Does the paper specify all the training and test details (e.g., data splits, hyperparameters, how they were chosen, type of optimizer) necessary to understand the results?
    \item[] Answer: \answerYes{}
    \item[] Justification: Appendix~\ref{app:training} reports the optimizer (Adam with default $\beta_1{=}0.9$, $\beta_2{=}0.999$), the batch-size scheduler, early stopping (patience 20), the train/validation/test splits used for each dataset, the hidden dimension per benchmark, the number of attention heads (4), the patch size ($P{=}10$), and the GMP grid spacing ($\delta=0.2$). Appendix~\ref{app:ablations} reports the hyperparameter sweep used to fix the patch-token configuration. Section~\ref{sec:results} states that all results are reported as mean $\pm$ standard deviation across $n{=}3$ random seeds.

\item {\bf Experiment statistical significance}
    \item[] Question: Does the paper report error bars suitably and correctly defined or other appropriate information about the statistical significance of the experiments?
    \item[] Answer: \answerYes{}
    \item[] Justification: All main-text and appendix tables report mean $\pm$ standard deviation over $n{=}3$ training runs with different random seeds; the variability captured is therefore initialization and stochastic optimization noise at fixed splits. The improvement of PHAT-JeT over JEDI-Linear in accuracy is additionally tested with a two-sided Welch $t$-test ($p{=}0.007$), reported in Section~\ref{sec:results}. ROC AUC uncertainties below $0.001$ are stated as such in the table captions rather than expanded.

\item {\bf Experiments compute resources}
    \item[] Question: For each experiment, does the paper provide sufficient information on the computer resources (type of compute workers, memory, time of execution) needed to reproduce the experiments?
    \item[] Answer: \answerYes{}
    \item[] Justification: Training is performed on a single NVIDIA RTX 3090 GPU; a full training run completes in approximately 5 hours. Inference and benchmarking are performed on a single NVIDIA GTX 1080 Ti. The proposed model is small ($\sim$7K parameters, $\sim$1.3\,M FLOPs per forward pass at $N{=}150$), so the total compute used in this work, including the hyperparameter sweep, ablations, and $n{=}3$ seed averaging across all four benchmarks, fits comfortably within typical jet-tagging research budgets. These details are also recorded in Appendix~\ref{app:training}.

\item {\bf Code of ethics}
    \item[] Question: Does the research conducted in the paper conform, in every respect, with the NeurIPS Code of Ethics \url{https://neurips.cc/public/EthicsGuidelines}?
    \item[] Answer: \answerYes{}
    \item[] Justification: The authors have reviewed the NeurIPS Code of Ethics and the research conforms with it. The work uses publicly available simulated jet-tagging benchmarks, does not involve human subjects or personal data, and does not raise dual-use concerns of the type the Code addresses.

\item {\bf Broader impacts}
    \item[] Question: Does the paper discuss both potential positive societal impacts and negative societal impacts of the work performed?
    \item[] Answer: \answerYes{}
    \item[] Justification: The Impact Statement in the main text discusses both the positive societal impact of more efficient real-time triggering at the LHC (improved scientific reach for fundamental physics measurements at fixed compute and energy budgets) and the limitations of training on simulated rather than measured collision data. We additionally note that the methods are domain-specific to high-energy physics and have no obvious dual-use or misuse pathway.

\item {\bf Safeguards}
    \item[] Question: Does the paper describe safeguards that have been put in place for responsible release of data or models that have a high risk for misuse (e.g., pre-trained language models, image generators, or scraped datasets)?
    \item[] Answer: \answerNA{}
    \item[] Justification: The released artifact is a small jet-tagging classifier ($\sim$7K parameters) trained on publicly available simulated collision data. It is purpose-built for classifying particle jets at the LHC and has no plausible misuse pathway analogous to language models, image generators, or scraped web datasets.

\item {\bf Licenses for existing assets}
    \item[] Question: Are the creators or original owners of assets (e.g., code, data, models), used in the paper, properly credited and are the license and terms of use explicitly mentioned and properly respected?
    \item[] Answer: \answerYes{}
    \item[] Justification: All four datasets used (\textsc{hls4ml}, JetClass, Top Tagging, Quark--Gluon) are cited at their first use and used under their respective public licenses. Baseline implementations referenced or adapted from prior work (JEDI-Linear, Linformer, SAL-T, PointNet, Point Transformer V3, HEPT) are cited where they appear in tables and in the text. We use these datasets and baselines for their intended research purpose (jet-tagging benchmarking) and respect the terms of use under which they were released.

\item {\bf New assets}
    \item[] Question: Are new assets introduced in the paper well documented and is the documentation provided alongside the assets?
    \item[] Answer: \answerYes{}
    \item[] Justification: The new asset introduced is the PHAT-JeT model and its training pipeline. The anonymized code release accompanying the submission includes the model definition, training and evaluation scripts, configuration files for each benchmark, and a README documenting environment setup, data preprocessing, and the commands to reproduce the main-text tables. No new datasets are released.

\item {\bf Crowdsourcing and research with human subjects}
    \item[] Question: For crowdsourcing experiments and research with human subjects, does the paper include the full text of instructions given to participants and screenshots, if applicable, as well as details about compensation (if any)?
    \item[] Answer: \answerNA{}
    \item[] Justification: This work does not involve crowdsourcing or research with human subjects. All training and evaluation data are simulated jet samples produced by standard particle-physics Monte Carlo generators.

\item {\bf Institutional review board (IRB) approvals or equivalent for research with human subjects}
    \item[] Question: Does the paper describe potential risks incurred by study participants, whether such risks were disclosed to the subjects, and whether Institutional Review Board (IRB) approvals (or an equivalent approval/review based on the requirements of your country or institution) were obtained?
    \item[] Answer: \answerNA{}
    \item[] Justification: This work does not involve human subjects, so no IRB approval was required.

\item {\bf Declaration of LLM usage}
    \item[] Question: Does the paper describe the usage of LLMs if it is an important, original, or non-standard component of the core methods in this research? Note that if the LLM is used only for writing, editing, or formatting purposes and does \emph{not} impact the core methodology, scientific rigor, or originality of the research, declaration is not required.
    %this research?
    \item[] Answer: \answerNA{}
    \item[] Justification: LLMs are not part of the core methodology of this work. PHAT-JeT is a from-scratch transformer architecture trained on jet-tagging benchmarks; no LLM is used as a component of the model, the training pipeline, or the evaluation. Any use of LLMs by the authors during manuscript preparation was limited to writing, editing, and formatting and did not affect the scientific content.

\end{enumerate}
\end{document}